\documentclass[12pt,dvips]{article}
\usepackage{graphicx}
\usepackage{amsmath}
\usepackage[psamsfonts]{amssymb}
\usepackage{amsthm}
\usepackage{euscript}

\usepackage{latexsym}

\renewcommand{\theequation}{\arabic{section}.\arabic{equation}}
\renewcommand{\(}{\begin{equation}}
\renewcommand{\)}{end{equation} \vspace{-.05in}\linebreak}

\newcounter{saveeqn}
\newcounter{savealpheqn}

\newcommand{\alpheqn}{\setcounter{saveeqn}{\value{equation}}%
 \stepcounter{saveeqn}\setcounter{equation}{0}%
 \renewcommand{\theequation}{\mbox{\arabic{section}.\arabic{saveeqn}\alph{equation}}}
 \renewcommand{\)}{\end{equation}}}
\def\part#1{\frac{\partial}{\partial{#1}}}%
\def\group#1{\refstepcounter{equation}\setcounter{saveeqn}{\value{equation}}%
 \label{#1}\setcounter{equation}{0}%
 \renewcommand{\theequation}{\mbox{\arabic{section}.\arabic{saveeqn}\alph{equation}}}
 \renewcommand{\)}{\end{equation}}}
\newcommand{\reseteqn}{\setcounter{equation}{\value{saveeqn}}%
 \renewcommand{\theequation}{\arabic{section}.\arabic{equation}}%
 \renewcommand{\)}{\end{equation}}}

\newcommand{\aalpheqn}{\setcounter{saveeqn}{\value{equation}}%
 \stepcounter{saveeqn}\setcounter{equation}{0}%
 \renewcommand{\theequation}{\mbox{\Alph{subsection}.\arabic{saveeqn}\alph{equation}}}
  \renewcommand{\)}{\end{equation}}}
\newcommand{\areseteqn}{\setcounter{equation}{\value{saveeqn}}%
 \renewcommand{\theequation}{\Alph{subsection}.\arabic{equation}}%
 \renewcommand{\)}{\end{equation}}}

\renewcommand{\thefootnote}{\alph{footnote}}
\renewcommand{\(}{\begin{equation}}
\renewcommand{\)}{\end{equation}}
\newcommand{\ba}{\begin{eqnarray}}
\newcommand{\ea}{\end{eqnarray}}

\newcommand{\bp}{\mathop{\vtop{\ialign{##\crcr
  $\hfil\displaystyle{}\hfil$\crcr\noalign{\kern-13pt\nointerlineskip}
  \BIG{(}\hskip0pt\crcr\noalign{\kern3pt}}}}}
\newcommand{\cbp}{\mathop{\vtop{\ialign{##\crcr
  $\hfil\displaystyle{}\hfil$\crcr\noalign{\kern-13pt\nointerlineskip}
  \BIG{)}\hskip0pt\crcr\noalign{\kern3pt}}}}}
\newcommand{\pa}{\mathop{\vtop{\ialign{##\crcr
  $\hfil\displaystyle{\oplus}\hfil$\crcr\noalign{\kern+1pt\nointerlineskip}
  \hspace{.08in}$^{\alpha=0}$\hskip6pt\crcr\noalign{\kern3pt}}}}}
\renewcommand{\sp}{,\hspace{.3in}}
\newcommand{\p}{^\prime}

\newcommand{\appendixa}
 {\renewcommand{\theequation}{\Alph{subsection}.\arabic{equation}}%
  \renewcommand{\thesubsection}%
               {Appendix \Alph{subsection}.\setcounter{equation}{0}}%
  \renewcommand{\alpheqn}{\aalpheqn}%
  \renewcommand{\reseteqn}{\areseteqn}
  }

\newcommand{\R}{\ensuremath{\mathbb R}}

\newcommand{\Z}{\ensuremath{\mathbb Z}}

\newcommand{\N}{\ensuremath{\mathcal N}}
\newcommand{\beq}{\begin{equation}}
\newcommand{\eeq}{\end{equation}}

\numberwithin{equation}{section}
\def\hsp#1{\hspace{#1in}}

\catcode`\@=11
\def\vereq#1#2{\lower3pt\vbox{\baselineskip1.5pt \lineskip1.5pt
\ialign{$\m@th#1\hfill##\hfil$\crcr#2\crcr\sim\crcr}}}
\catcode`\@=12

\makeatletter
\newcommand\figcaption{\def\@captype{figure}\caption}
\newcommand\tabcaption{\def\@captype{table}\caption}
\makeatother

\begin{document}
\begin{titlepage}
\begin{flushright}
UCB-PTH-01/24 \\
hep-th/0107072
\end{flushright}

\vspace{2em}
\def\thefootnote{\fnsymbol{footnote}}

\begin{center}
{\Huge Dial M for Flavor Symmetry Breaking}
\end{center}
\vspace{1em}

\begin{center}
Jarah Evslin\footnote{E-Mail: jarah@uclink4.berkeley.edu}, Hitoshi Murayama\footnote{E-Mail: murayama@lbl.gov}, Uday Varadarajan\footnote{E-Mail: udayv@socrates.berkeley.edu} and John E. Wang\footnote{E-Mail:
  hllywd2@physics.berkeley.edu}
\end{center}

\begin{center}
\vspace{1em}
{\em Department of Physics,
     University of California\\
     Berkeley, California 94720}\\
and\\
{\em Theoretical Physics Group\\
     Ernest Orlando Lawrence Berkeley National Laboratory\\
     University of California,
     Berkeley, California 94720}        

\end{center}

\vspace{3em}
\begin{abstract}

\noindent
We introduce a realization of Seiberg duality in MQCD that does not
involve moving branes. Using related arguments in M Theory and IIA,
we investigate the flavor symmetry breaking via flavored magnetic
monopole condensation of hep-th/0005076. We verify our results by
considering visualizations of M5 branes corresponding to $\N=2$ vacua
which survive soft breaking by an adjoint mass term.

\end{abstract}

\vfill
August 8, 2001

\end{titlepage}
\setcounter{footnote}{0} 
\renewcommand{\thefootnote}{\arabic{footnote}}

\pagebreak
\renewcommand{\thepage}{\arabic{page}}
\pagebreak 

\section{Introduction}

Four years ago we learned that superQCD is contained in the low energy
description of certain configurations of M5 branes in multi-centered
Taub-NUT space \cite{Witten}.  As a result, brane configurations in M
theory are able to provide a concrete realization of otherwise
mysterious quantum phenomena, such as confinement via magnetic
monopole condensation \cite{SeibergWitten1,Witten2,Rabinovici} and
Douglas-Shenker strings \cite{DS,HananyZaffaroni}.  This approach is
known as MQCD and is reviewed in \cite{Giveon}. 

In this paper, we introduce a novel MQCD description of Seiberg
duality in $\N=2$ superQCD softly broken to $\N=1$ by adding a
mass term for the adjoint chiral multiplets. We then proceed to
construct the M theory realization of the U($N_f$) $\rightarrow$
U($r$)$\times$U($N_f-r$) flavor symmetry breaking mechanism in $\N=1$
SU($N_c$) gauge theories studied in Refs.~\cite{Murayama1, Murayama,
  Murayamatalk}.

In section~\ref{revsec}, we will review the semi-classical description
of supersymmetric gauge theories as effective field theories of
parallel D4 branes suspended between NS5 branes in type IIA string
theory.  To understand the quantum gauge theories we lift this
description to M theory.  We then review the field theory results of
Refs.~\cite{Murayama1, Murayama, Murayamatalk} in section 3.  Then, in
section 4 we describe Seiberg duality as a choice of two D brane
configurations in IIA which lift to the same M5 brane configuration as
seen by M2 brane probes at different energy scales.  We finally turn
to the MQCD description of flavor symmetry breaking in section 5. A
detailed example of the M5 brane configurations corresponding to
r-vacua in SU(3) with 4 flavors is presented in Appendix A.

\section{A Review of MQCD} \label{revsec}

We begin with a review of the standard embedding of $3+1$ dimensional
$SU(N_c)$ SQCD in IIA string theory and its M theory lift.

\subsection{Classical $\mathcal{N}=2$ SQCD and IIA}

\begin{figure}[ht]
  \centering \includegraphics[width=6in]{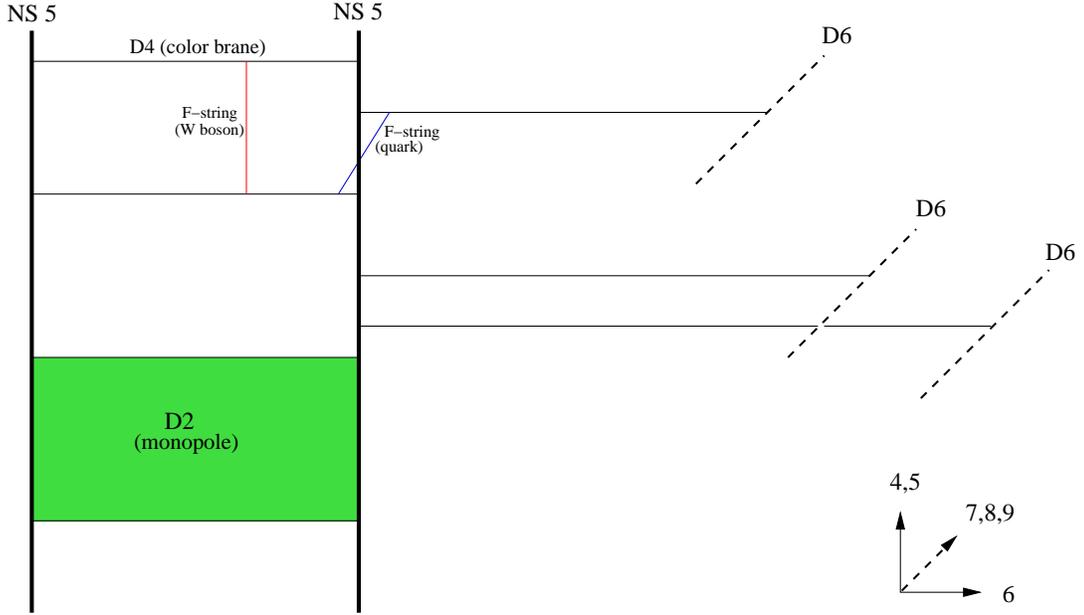}
\caption{Type IIA string theory realization of $\N=2$ SU(4) SQCD with
  3 flavors. The effective worldvolume of the gauge theory (which lives on the
  D4 color branes) is spanned by the $x^0$, $x^1$, $x^2$, and $x^3$
  directions.}
  \label{basic}
\end{figure}

Consider type IIA string theory on $\R^{9,1}$ with coordinates $x^0,
\ldots, x^9$ and complex coordinates
\begin{equation}
v=x^4+i x^5\sp w=x^8+ix^9. 
\end{equation}
A stack of $N_c$ parallel D4 ``color'' branes extend along directions
$x^0$, $x^1$, $x^2$, $x^3$, and $x^6$.  The low energy theory on these
branes is SU($N_c$) 4+1 dimensional super Yang-Mills with 16
supercharges.  We can Kaluza-Klein reduce this to the desired 3+1
dimensional SYM with 8 supercharges by adding two parallel NS5 branes
extended along the $x^0$, $x^1$, $x^2$, $x^3$, $x^4$, and $x^5$
directions and placed at positions $0$ and $L_6$ along the $x^6$
direction with the color branes suspended between them.

The effective gauge coupling, $g$, of the 3+1 dimensional theory is given by 
\begin{equation}
\frac{1}{g^2}=\frac{L_6}{g_s l_s}
\end{equation}
where $g_s$, $l_s$, and $L_6$ are the string coupling constant, the
string length and the distance between the two NS5 branes.  To
decouple the degrees of freedom of the bulk from the
color branes we take the limits
\begin{equation}
g_s \ \rightarrow \ 0 \sp \frac{L_6}{l_s} \ \rightarrow \ 0 \sp g=\textup{constant}.
\end{equation}
The light perturbative degrees of freedom of the gauge theory are
strings which stretch between the color branes, yielding an
$\mathcal{N}=2$ vector multiplet transforming in the adjoint
representation of SU($N_c$).  The distances between color branes
correspond to the vacuum expectation values of the adjoint scalars in
the vector multiplet and so parameterize the Coulomb branch. 



Quark hypermultiplets transforming in the fundamental representation
of SU($N_c$) may be included by attaching $N_f$ D4 ``flavor'' branes
stretching between one of the NS5 branes and a D6 flavor brane.  To
preserve $\N=2$ supersymmetry no two D4 branes may connect the same
NS5 and D6 brane.  This is known as the s-rule {\cite{HananyWitten}}
and is U-dual {\cite{Bachas1, Bachas2}} to Pauli's exclusion
principle. The quarks and squarks are strings which stretch from one
D4 flavor brane to one D4 color brane and so they transform in the
fundamental representations of both the SU($N_c$) gauge group and the
global flavor symmetry group. Semiclassical magnetic monopole and dyon
{\cite{Witten, HY, Russian}} states are realized by D2 branes with the
topology of a disk bounded by a D4-NS5-D4-NS5 cycle. The brane
configuration is summarized in Table~\ref{bral}.

\begin{figure}[h]
\centering
\begin{tabular}{c|c|c|c|c|c|c|c|c|c|c}
Brane & $x^0$ & $x^1$ & $x^2$ & $x^3$ & $x^4$ & $x^5$ & $x^6$ & $x^7$
& $x^8$ & $x^9$ \\ \hline
D4           & X & X & X & X &     &     & X &    &     &     \\
NS5          & X & X & X & X &  X  &  X  &   &    &     &     \\
NS5$_\theta$ & X & X & X & X & (X) & (X) &   &    & (X) & (X) \\
D6           & X & X & X & X &     &     &   & X  &  X  &  X  \\
D2           & X &   &   &   & (X) & (X) & X &    &     &     \\
D0           & (X) &   &   &   &     &     & (X) &    &     &     \\
\multicolumn{10}{c}{}\\
\end{tabular}
\tabcaption{Alignments of branes in IIA. Parentheses indicate that the
brane may be aligned at an angle between the given directions. For the
D0 branes they are used to indicate the presence of both
D-instantons and dynamical D0 particles.} \label{bral}
\end{figure}

There is an unbroken global U($N_f$) symmetry when the flavor branes
are placed at the same $v$ coordinate, although this symmetry is
broken to U(1$)^{N_f}$ when they are placed at distinct positions
$v=m_i$, $i=1, \ldots, N_f$.  The $m_i$ are the bare quark masses.
Generally, a quark with flavor $i$ and color $a=1, \ldots,
N_c$ has mass
\begin{equation}
m_i^a=|m_i - \phi^a|
\end{equation}
which is the shortest distance between color brane $a$ and flavor
brane $i$.  Alternately the quark mass can be read from the
superpotential terms:
\begin{equation}
W \supset \int d^2 \theta \ \{ {\tilde{Q}}_i \Phi Q^i + m_i
{\tilde{Q}}_i Q^i \} 
\end{equation}
where $Q_i, {\tilde{Q}}_i$ are the $\N=1$ chiral multiplets of the
quark hypermultiplet and $\Phi$ is the $\N=1$ chiral multiplet of the
$\N=2$ vector multiplet.

\begin{figure}[htb]
\centering
\includegraphics[width=6in]{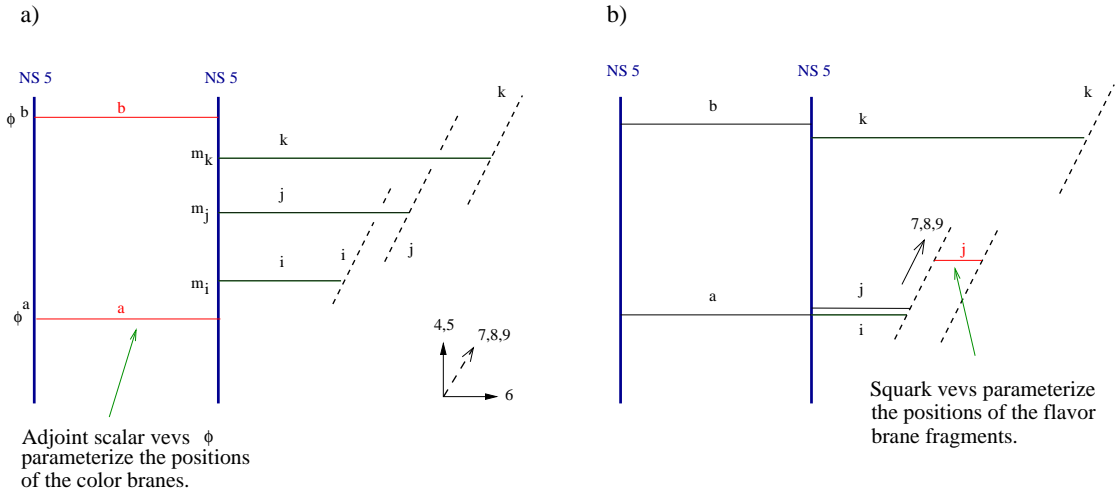} 
\hsp{-1}\caption{Type IIA brane configurations corresponding to the a)
  Coulomb branch b) a mixed Coulomb and Higgs branch.}
\label{quark}
\end{figure}

When two flavor branes $i$ and $j$ and a color brane $a$ are at the
same $v$ coordinate, $\phi^a=m_i=m_j$, then it is possible to enter
the Higgs branch of the gauge theory.  This is done by connecting
color brane $a$ to flavor brane $i$ and then breaking flavor brane $j$
on D6 brane $i$.  At this point we are allowed to move the portion of
D4 brane $j$ which is between D6 branes $i$ and $j$, corresponding to
generating vacuum expectation values for the squarks in the
hypermultiplet.  These vacuum expectation values are parameterized by
the position of the D4 brane in the $x^7$, $x^8$, and $x^9$ directions as
well as the Wilson line of the gauge field $A_6$.


The U(1$)_R$ $\times$ SU(2$)_R$ R-symmetry of the classical $\N=2$
theory is manifested as a rotational symmetry of the brane cartoon.
The U(1$)_R$ symmetry corresponds to rotations of the $v$-plane, while
the SU(2$)_R$ is the universal cover of the SO(3) acting on $x^7$,
$x^8$, and $x^9$ by rotations.

\subsection{Quantum Gauge Theory and M Theory}

By lifting the above brane configuration to M theory {\cite{Witten}},
we can consider non-perturbative (in $g_s$) quantum corrections and
thereby gain considerable insight into the origin of quantum phenomena
in SQCD. In particular, we consider the above IIA brane configuration
in the limit of large $g_s$ and $L_6$ with $\frac{1}{g^2}=\frac{L_6}{R}$
fixed, where the classical description of M theory is valid.

In M theory, the D4 brane is an M5 brane which wraps the M theory
circle ($x^{10}\sim x^{10}+2\pi$) once while an NS5 brane is an M5
brane which does not wrap the $x^{10}$ direction. The collection of D4
and NS5 branes can therefore be described as a single M5 brane
$\R^{3,1} \times \Sigma$ which fills the $x^0$, $x^1$, $x^2$, $x^3$
space and is a Riemann surface $\Sigma$ in the $x^4$, $x^5$, $x^6$,
and $x^{10}$ directions.

\begin{figure}[htb]
\centering
\includegraphics[width=4in]{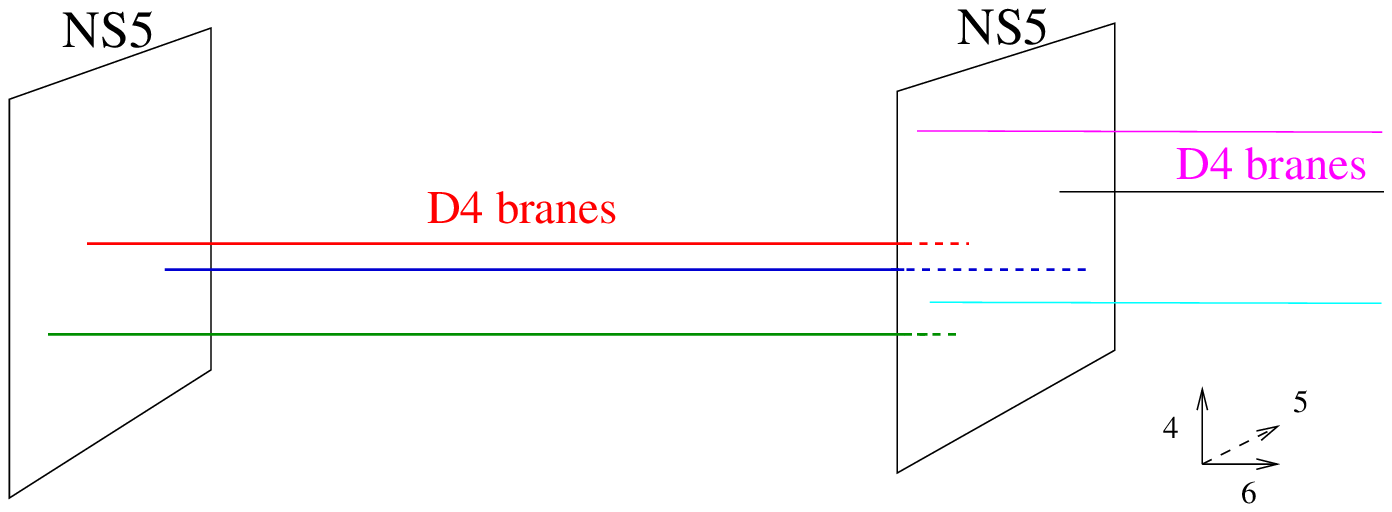}
\caption{IIA realization of the Coulomb branch of $\N=2$ SU(3) SQCD
  with 3 flavors} \label{m5}
\includegraphics[width=4in]{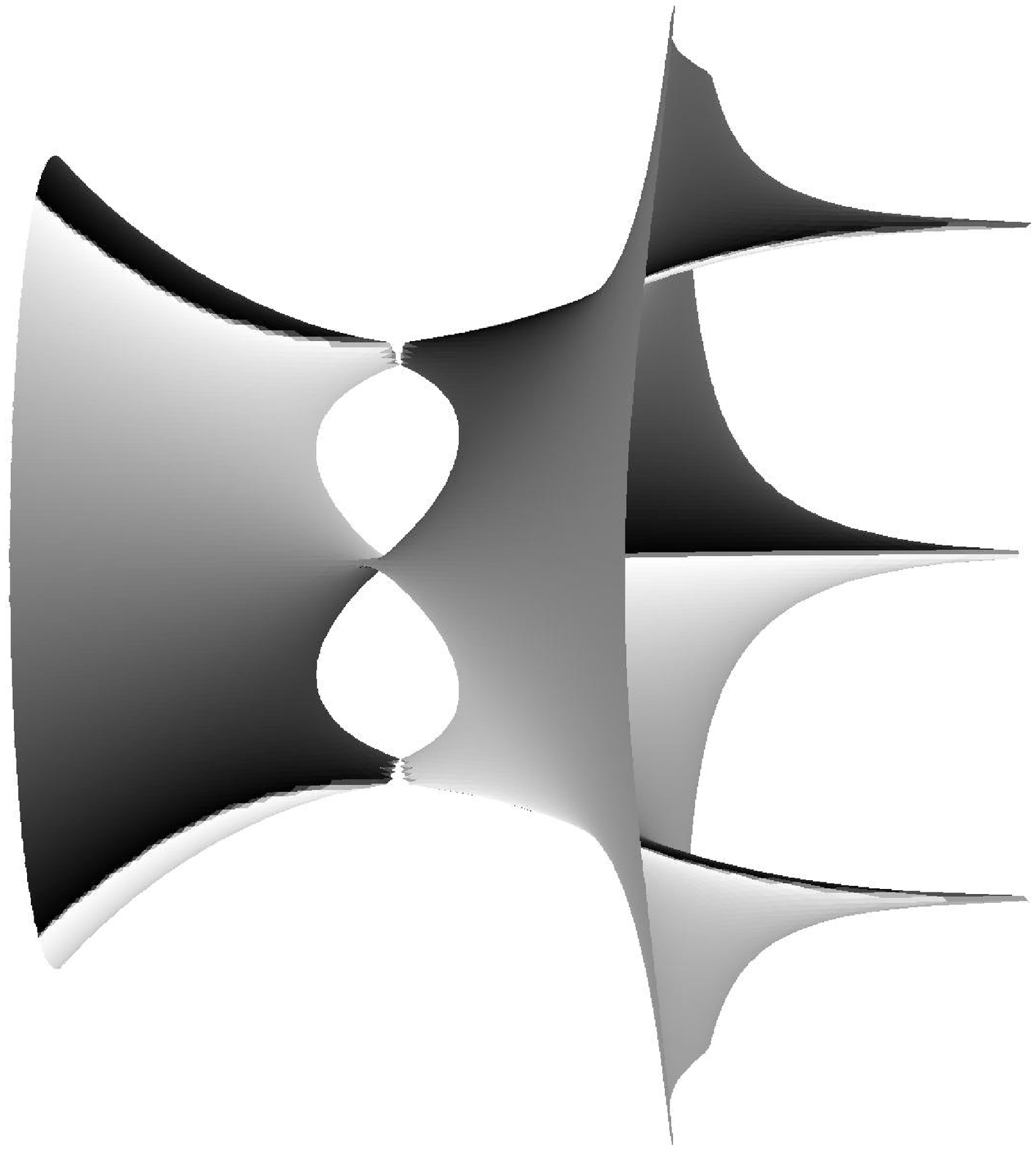}
\caption{The M theory lift of the above IIA configuration to a single
  M5 brane. The directions $x^4$, $x^5$, and $x^6$ are shown explicitly
  while the $x^{10}$ coordinate is parameterized by darkness.}
  \label{m5lift}
\end{figure}

If we introduce the new holomorphic coordinate,
\begin{equation}
t=\exp{\left( \frac{-x^6}{R}-i x^{10} \right)}
\end{equation}
where $R=g_s l_s$ is the radius of the M theory circle, we can construct
$\Sigma$ explicitly as the vanishing locus of a polynomial $F(v,t)$,
\begin{equation}
F(v,t)=t^2+t\prod_{a=1}^{N_c} (v-\phi^a)+\Lambda^{2N_c-N_f}\prod_{i=1}^{N_f}(v-m_i)=0.
\end{equation}
where $\Lambda$ is the dynamically generated QCD scale of the theory.
Note that $\Sigma$ is precisely the Seiberg-Witten curve
\cite{SeibergWitten1} of the gauge theory. For example, the lift of
the brane configuration shown in Fig.~\ref{m5} is the single M5 brane
pictured in Fig.~\ref{m5lift}.


We can immediately see the running of the gauge coupling from the form
of the curve. Since the couplings are functions of the mass scales of
the lightest charged states in the theory, and as these scales just
correspond to distances in $v$ {\cite{Witten}}, we find that
\begin{equation} \label{grun}
\frac{1}{g(v)^2}=\frac{L_6(v)}{g_s l_s} \ \sim \ \log |v| \ .
\end{equation}
Also, the U(1$)_R$ anomaly is particularly easy to see in the M theory
picture. Recall that D4 branes wrap the M theory direction and end on
the NS5 branes, which do not wrap the M theory direction.  This means
that the end of a D4 brane on an NS5 brane is a vortex in the
embedding coordinate of the NS5 brane in the M theory direction (see
Figs.~\ref{m5lift} and \ref{Uday}). In particular, at large $v$ if one
follows a circle along the NS5 brane whose interior contains all of
the D4 branes, this circle will wrap the M theory direction as many
times as there are colors minus flavors attached to this NS5 brane.
Thus the naive U(1$)_R$ rotational R-symmetry of the brane must be
combined with a simultaneous rotation of the M theory circle. Such a
redefinition is not possible with both NS5 branes as the $x^{10}$
redefinitions would have to be in opposite directions for the two
branes and so the U(1$)_R$ is anomalous. There is a residual
$\Z_{2N_c-N_f}$ symmetry in the U(1) redefined to include a rotation
of the M theory circle in opposite directions on both NS5 branes. For
these special angles the two redefinitions only disagree by a multiple
of $2\pi$ in the M theory direction.

The reader may verify these claims visually by considering the
asymptotic $x^{10}$ dependence of the regions corresponding to the NS5
branes (large positive and negative values of $x^6$) on the M5 branes
pictured in Figs.~\ref{m5lift} and \ref{Uday}.

Another benefit of the M theory lift is that all the matter in the
gauge theory is realized in M theory by open M2 branes ending on the
M5 brane. In reducing to IIA, the M2 branes which wrap the M theory
circle carry fundamental string charge, while unwrapped M2 branes are
D2 branes. The M2 brane configurations corresponding to BPS states
were analyzed critically in \cite{HY,Russian}. They found that BPS
states were M2 branes with the minimal possible areas in their
homology classes.

\begin{figure}[htb]
\centering
\includegraphics[width=6in]{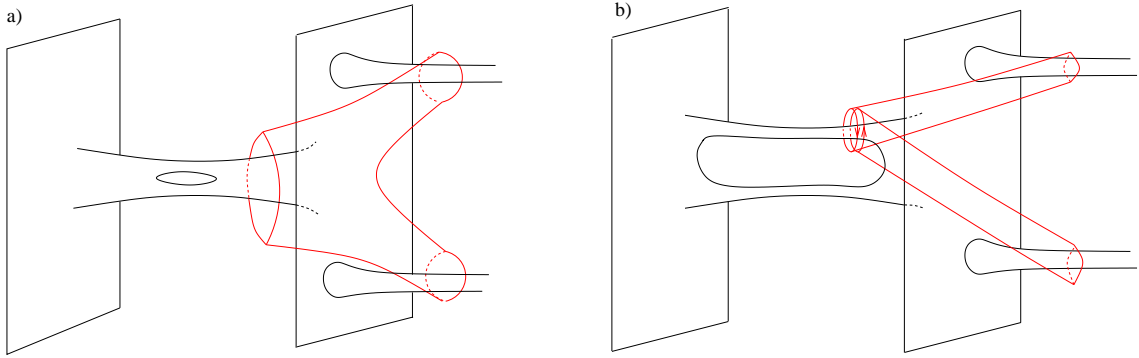}
\caption{a) a baryon which wraps 2 color and 2 flavor branes in SU(2)
  SQCD with 2 flavors\hsp{.5} {}b) a meson which wraps 2 flavor branes
  and wraps and unwraps one color brane} \label{baryon}
\end{figure}

We can also gain intuition for baryons and mesons in SQCD using M2
branes. For example, baryons are M2 branes {\cite{Rabinovici}} with
$k+1$ boundaries as seen in Figure~\ref{baryon}.  One boundary wraps
$k$ color branes and the other $k$ boundaries each wrap a flavor
brane.  In the IIA limit, the baryon is a collection of $k$ quarks
attached by a k-string {\cite{Witten2,HananyZaffaroni}}.  Mesons are
tubular M2 branes which start by wrapping one flavor brane, connect to
a color brane, and then extend in the space directions ($x^1$, $x^2$,
and $x^3$) before wrapping another flavor brane.

We certainly have gained a great deal in this lift, but we don't quite
get it for free. Since this lift involves going to large string
coupling, one cannot be sure that the low energy effective theory on
the M5 brane is the gauge theory we started with. In fact, it is
actually a six dimensional theory. The M theory limit is exactly the
opposite limit needed to obtain the 4D gauge theory. It requires
taking the radius of the M theory circle $R=g_s l_s$ as well as $L_6$
large leaving $\frac{1}{g^2}=\frac{L_6}{R}$ fixed.  However, since
unbroken supersymmetries protect certain holomorphic quantities (like
the masses of BPS states and superpotentials) from perturbative $g_s$
quantum corrections, one can still use classical computations
involving the M5 brane to determine them exactly.  The breakdown of
the M theory picture for the computation of non-holomorphic quantities
(like the masses of non-BPS states) can then be understood as due to
the presence of KK modes (dynamical D0 branes and $A_6$ fluctuations)
which become light and strongly coupled in this limit \cite{Witten2}.

For example, one can solve the $\N=2$ gauge theories constructed above
at generic points in their moduli spaces as their effective actions
are governed by holomorphic quantities ($\N=2$ prepotentials ${\cal
  F}$) which can be computed exactly using $\Sigma$ {\cite{Witten}}.
Upon breaking to $\N=1$, the K\"{a}hler potential is no longer
protected from such corrections. However, as the superpotential of the
$\N=1$ theory is still protected, we will see that we can still learn
what we need about $\N=1$ SQCD from M theory.

\subsection{Soft Breaking to $\mathcal{N}=1$ SQCD} {\label{N1section}}

Rotating one of the NS5 branes
\begin{equation} 
\left( \begin{array}{c} v\p \\  w\p \end{array} \right) = \left(
  \begin{array}{cc} \cos \theta & \sin\theta \\  -\sin \theta & \cos
    \theta \end{array} \right) \left( \begin{array}{c} v \\  w
  \end{array} \right) \ 
\end{equation} 
results in a brane configuration which preserves only four
supercharges {\cite{Berkooz}} and softly breaks ${\mathcal{N}=2}$
supersymmetry to ${\mathcal{N}=1}$.  We will refer to the rotated NS5
brane as the NS$5_\theta$ brane. The NS$5_{\pi/2}$ brane is commonly
referred to as the NS$5\p$ brane in the literature.

\begin{figure}[htb]
\centering
\includegraphics[width=6in]{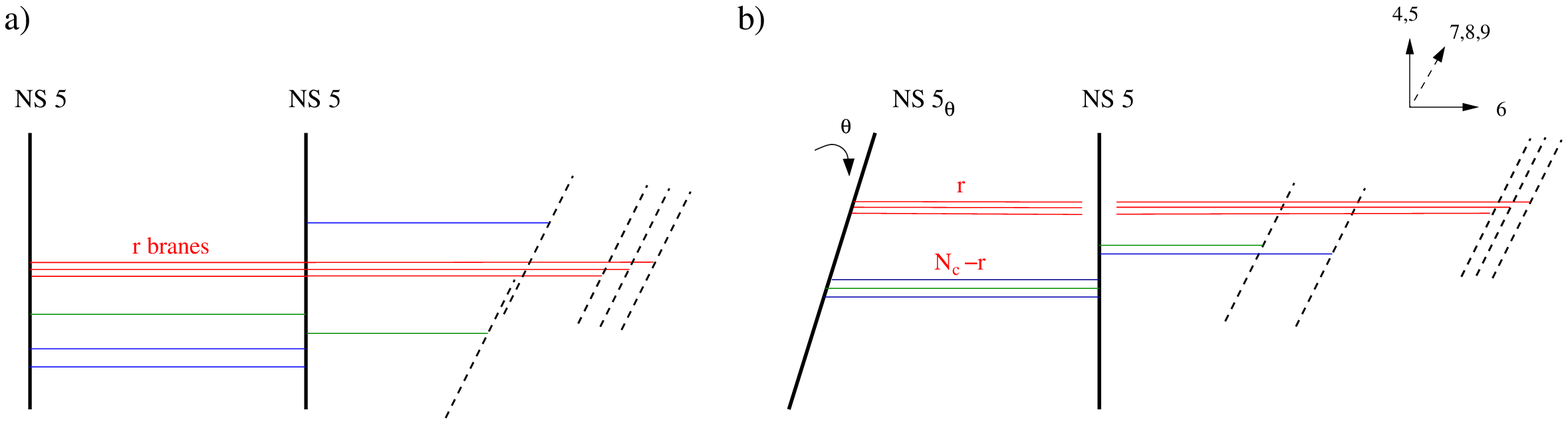}
\caption{a) $r$ color branes and flavor brane connects.  b) After
  rotating the five brane, color branes which are not connected to
  flavor branes move to the origin of $v$.  This is the realization of
  the $r$ vacua in \cite{Plesser}.} \label{3tflavor}
\end{figure}

\begin{figure}[htb]
\centering
\includegraphics[width=3in]{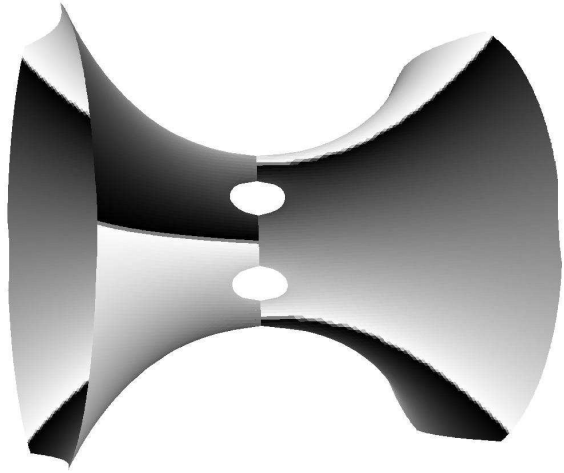}
\caption{An M5 brane configuration corresponding to $\N=2$ SU(3) SYM
  near a vacuum which survives upon softly breaking to $\N=1$. The two
  small holes are the degenerating cycles. The M2 branes corresponding
  to nearly massless monopoles are disks bounded by these cycles. The
  M theory direction is parameterized by darkness.}
\label{Uday}
\end{figure}

The symmetry breaking classical superpotential generated by this
process can be understood as follows. After the rotation, color branes
at generic positions in $v$ no longer minimize their lengths. In fact,
to reach equilibrium, all color branes much either slide to $v\sim
O(\Lambda)$ or attach to flavor branes.  Since translations along $v$
(which correspond to adjoint scalar vevs) now cause the color branes
to stretch, the $\N=1$ chiral multiplets containing these adjoint
scalars acquire a mass $\mu$ via the superpotential term
\begin{equation}
W\supset\mu \ \textup{Tr} \Phi^2\sp 
\mu \ \sim \ \tan \theta \ .
\end{equation}
If $r$ color branes connect to flavor branes, classically the only
surviving vacuum is
\begin{equation}
\phi^i=m_i \ \mbox{  for  } \ i=1, \ldots, r\hsp{.05} \hsp{.666}\phi^a=0
 \ \mbox{  for  } \ a=r+1, \ldots N_c.
\end{equation}
However, when quantum corrections are considered the surviving vacua
are those where $N_c-r$ or $N_c-r-1$ monopoles and dyons become
massless and condense with vevs of order $O(\mu\Lambda)$. In the M
theory picture, this occurs when the bounding cycles of the
corresponding M2 branes degenerate, as in Fig.~{\ref{Uday}}.

More precisely, the vacua which survive the $\mu \mbox{Tr} \Phi^{2}$
perturbation are those in which the cycles that have degenerated
result in an M5 brane configuration $\Sigma \times \R^4$ such that
$\Sigma$ is of genus zero, as argued in Refs.~{\cite{SeibergWitten1,
    Witten2, HoriOz}}. 

To preserve SUSY the flavor branes must continue to extend along the
$x^0$, $x^1$, $x^2$, $x^3$, and $x^6$ directions, in particular they
cannot rotate into the $w$ directions.  Therefore when the
NS$5_\theta$ brane rotates, the flavor branes ending on it must
translate in $w$, sliding along the corresponding D6 brane.  This
translation corresponds to meson vevs \cite{Giveon,HoriOz}.

\section{Summary of Field Theory Results} \label{sumfield}

The dynamics of the $\N=2$ supersymmetric SU($N_{c}$) gauge theories
constructed above and the dynamical breaking of flavor symmetry have
been studied in detail in Refs.~{\cite{Murayama1, Murayama,
Murayamatalk}}.  Here we briefly summarize the results.

The theory with $N_{f}$ massless quark hypermultiplets has U($N_{f}$) flavor
symmetry, SU(2$)_{R}$ symmetry, and a non-anomalous discrete
$\Z_{2N_{c}-N_{f}}$ subgroup of U(1$)_{R}$.  We are interested in the
$\N=1$ perturbation of the theory by the adjoint mass term $\mu
 \mbox{ Tr }\Phi^{2}$.  The moduli space contracts to the set of points
that give maximally degenerate (genus zero) Seiberg--Witten curves.

In the semi-classical regime with a large adjoint VEV, there are
`t~Hooft--Polyakov magnetic monopoles.  Zero modes of the quarks
around a monopole generate flavor quantum numbers for the magnetic
monopoles.  It was shown that they come in completely anti-symmetric
rank-$r$ tensor representations with $_{N_{f}}C_{r}$ multiplicities.

The strongly coupled regime was studied with a variety of techniques.
When $N_{f} < N_{c}$, there are vacua parameterized by an integer $r =
0, 1, \cdots, [N_{f}/2]$ ($[x]$ is the Gauss' symbol) where the flavor
symmetry is broken dynamically as U($N_{f}$) $\rightarrow$ U($r$)
$\times$ U($N_{f}-r$). If $r<N_f/2$, the physics around the vacuum can
be described by an IR free effective Lagrangian \cite{Plesser} (a
``magnetic dual'' to the asymptotically free semiclassical SU($N_c$)
description) with SU($r$) $\times$ U(1$)^{N_{c}-r-1}$ gauge group.
$N_{f}$ ``magnetic quark'' hypermultiplets transform as the
fundamental representation of SU($r$) while there are ``magnetic
monopoles'' for each of the ``magnetic'' U(1) factors.  When perturbed
by the adjoint mass term, all gauge groups are Higgsed by the
condensates of magnetic objects, corresponding to the confinement of
the electric theory.  It was argued in Ref.~\cite{Murayama1, Murayama}
that the semi-classical monopoles in the rank-$r$ anti-symmetric
tensor representation smoothly match to the baryonic composites of
magnetic quarks of the low-energy SU($r$) theory based on
circumstantial evidence.  Upon mass perturbations, one can count the
number of vacua:
\begin{equation}
    \N_{1} = (2N_{c}-N_{f})2^{N_{f}-1},
\end{equation}
originating from $r$-vacua ($r \leq [N_{f}/2]$) with $(2N_{c}-N_{f})$
copies due to the $\Z_{2N_{c}-N_{f}}$ symmetry.  Therefore {\it the
  flavor symmetry breaking and confinement have a common origin in
  these theories}\/: condensation of magnetic objects with non-trivial
flavor quantum numbers.  Strictly speaking, however, the existence of
monopoles in the anti-symmetric tensor representations was
demonstrated only in the semi-classical regime and its extrapolation
to the strongly coupled regime and the matching to the baryonic
composite was a conjecture.  When $r=N_{f}/2$ (possible obviously only
when $N_{f}$ is even), the low-energy magnetic gauge group is
superconformal with an infinitely strong coupling $\tau = -1$.  Due to
some reason, the same low-energy effective action seems to describe
the dynamics of flavor symmetry breaking even though there is no
weakly coupled description of the theory.

When $N_{f} > N_{c}$, there is a new vacuum without flavor symmetry
breaking.  It is at the same point on the moduli space as the
$r=N_{f}-N_{c}$ vacuum, while the finite quark mass perturbation shows
that there are additional
\begin{equation}
    \N_{2} = \sum_{r=0}^{N_{f}-N_{c}-1} (N_{f}-N_{c}-r)
    {}_{N_{f}}C_{r}
\end{equation}
vacua with unbroken flavor symmetry. 

\section{Seiberg Duality from M Theory} \label{MSeiberg}

As we noted in the last section, the effective theories at the
$r$-vacua which survive SUSY breaking are actually IR free theories
for $r < N_f/2$. In particular, at the baryonic root we find a weakly
coupled theory whose low energy physics is well described by an IR
free SU($\tilde{N}_c$) gauge theory, where $\tilde{N}_c = N_f - N_c$.
This is very reminiscent of Seiberg duality in $\N=1$ theories
\cite{Seiberg1}. In \cite{Plesser}, this observation was used to
``derive'' $\N=1$ Seiberg duality by mass perturbing the $\N=2$
theory\footnote{Unfortunately, the result was not quite Seiberg
  duality as there was an extra non-renormalizable coupling in the
  effective action associated with integrating out the massive adjoint
  scalar which becomes relevant in the low energy limit.}.  We will
consider this scenario using M theory.

\begin{figure}[htb]
\centering
\includegraphics[width=6in]{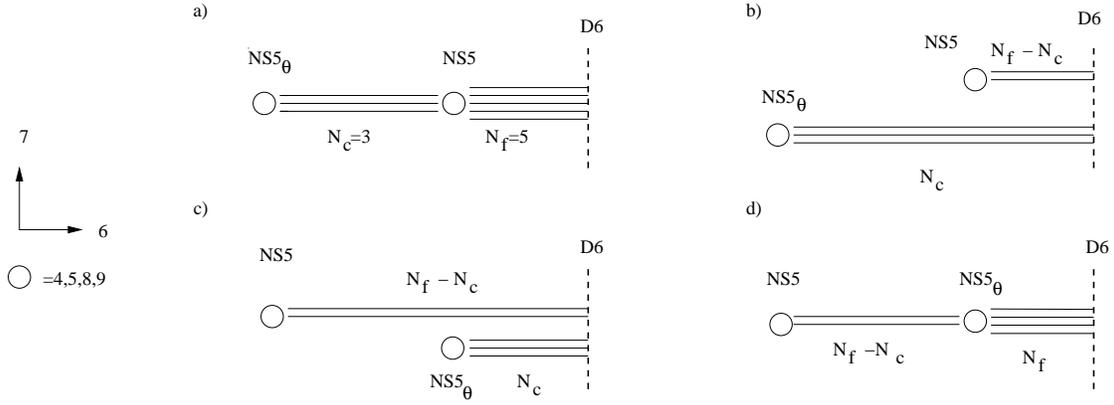}
\caption{In IIA Seiberg duality is realized by brane exchange.}
\label{sduality}
\end{figure}

Seiberg duality was first realized in IIA string theory in
Ref.~{\cite{Elitzur}} via the exchange of two NS5 branes, as
illustrated in Fig.~\ref{sduality}.  To avoid a singular configuration
the authors first displaced one of the NS5 branes in the 7 direction,
which corresponds to turning on a Fayet-Illipolous term in the field
theory description.  This process corresponds to Seiberg duality with
one caveat, the full U($N_f$) $\times$ U($N_f$) flavor symmetry is not
realized in the above IIA configuration nor in its M theory lift.

An alternate proposal for realizing Seiberg duality via string/M
theory was made later that year by Schmaltz and Sundrum \cite{SS}.
Using the M theory lift of the above brane setup, they found that
Seiberg duality could be understood by taking the $\Lambda\rightarrow
0$ and $\Lambda\rightarrow\infty$ limits of the resulting single M5
brane configuration. In particular, they found that the two limits
were related by an exchange of branes and therefore correspond to an
electric theory and its magnetic dual. However, their arguments
depended on turning on finite bare quark masses.

The following year this scenario was clarified and extended to the
massless case by Hori {\cite{Kentaro}}.  He observed that in M-theory,
the M5 branes can be crossed with no singularity even without the
Fayet-Illiopolous parameter.  In fact, at the root of the baryonic
branch, when $N_f \geq N_c+1$ the M5 brane consists of two connected
components \cite{HoriOz} and the duality corresponds to simply
translating one past the other along $x^6$.  The $x^6$ coordinate of a
connected component does not affect the field theory description and
so Hori argues that this process clearly preserves its universality
class.

Most recently, in Ref.~\cite{Dasgupta} Seiberg duality was understood
within the context of geometric engineering as a birational flop in a
T dual description consisting of D5 branes which are dual to the D4
branes in IIA and degenerations in the T-dualized circle which are
dual to the NS5 branes.

\begin{figure}[htb]
\centering
\includegraphics[width=6in]{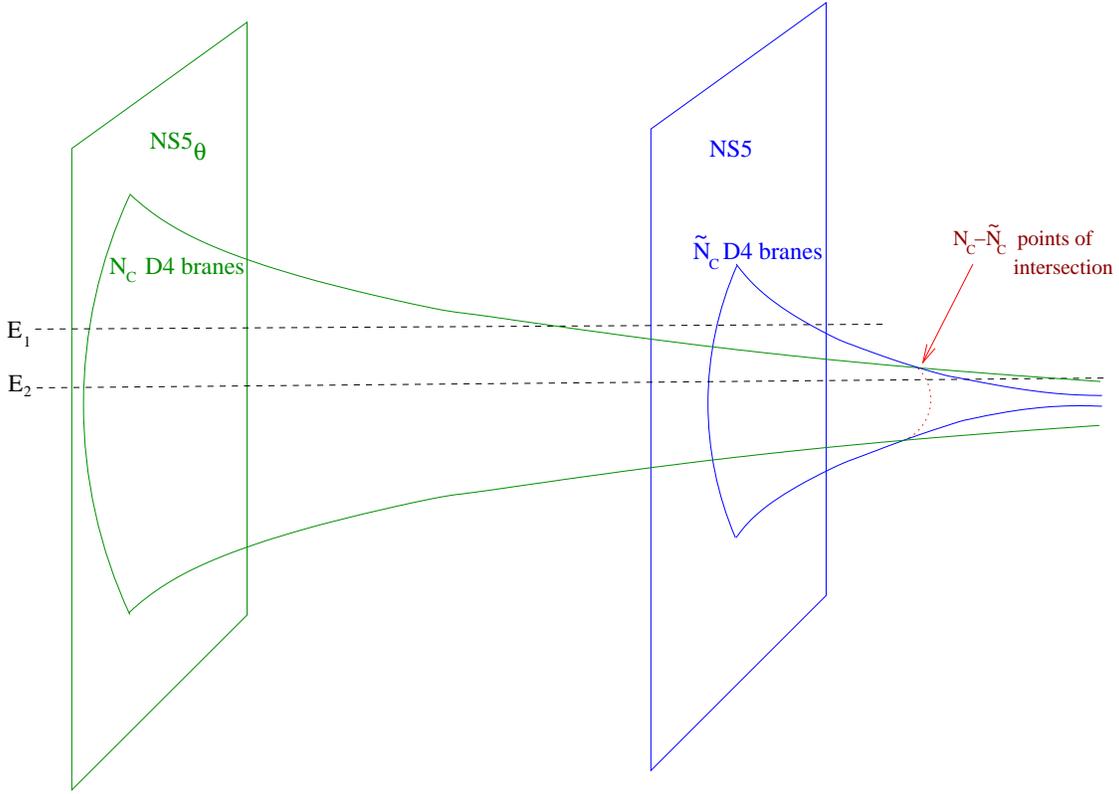}
\caption{At the root of the baryonic branch, the low energy physics
  (to the right of the intersection) is the SU($\tilde{N}_c$) $\times$
  U(1$)^{N_c-\tilde{N}_c}$ magnetic theory.} \label{duality}
\end{figure}

\begin{figure}[htb]
\centering
\includegraphics[width=4in]{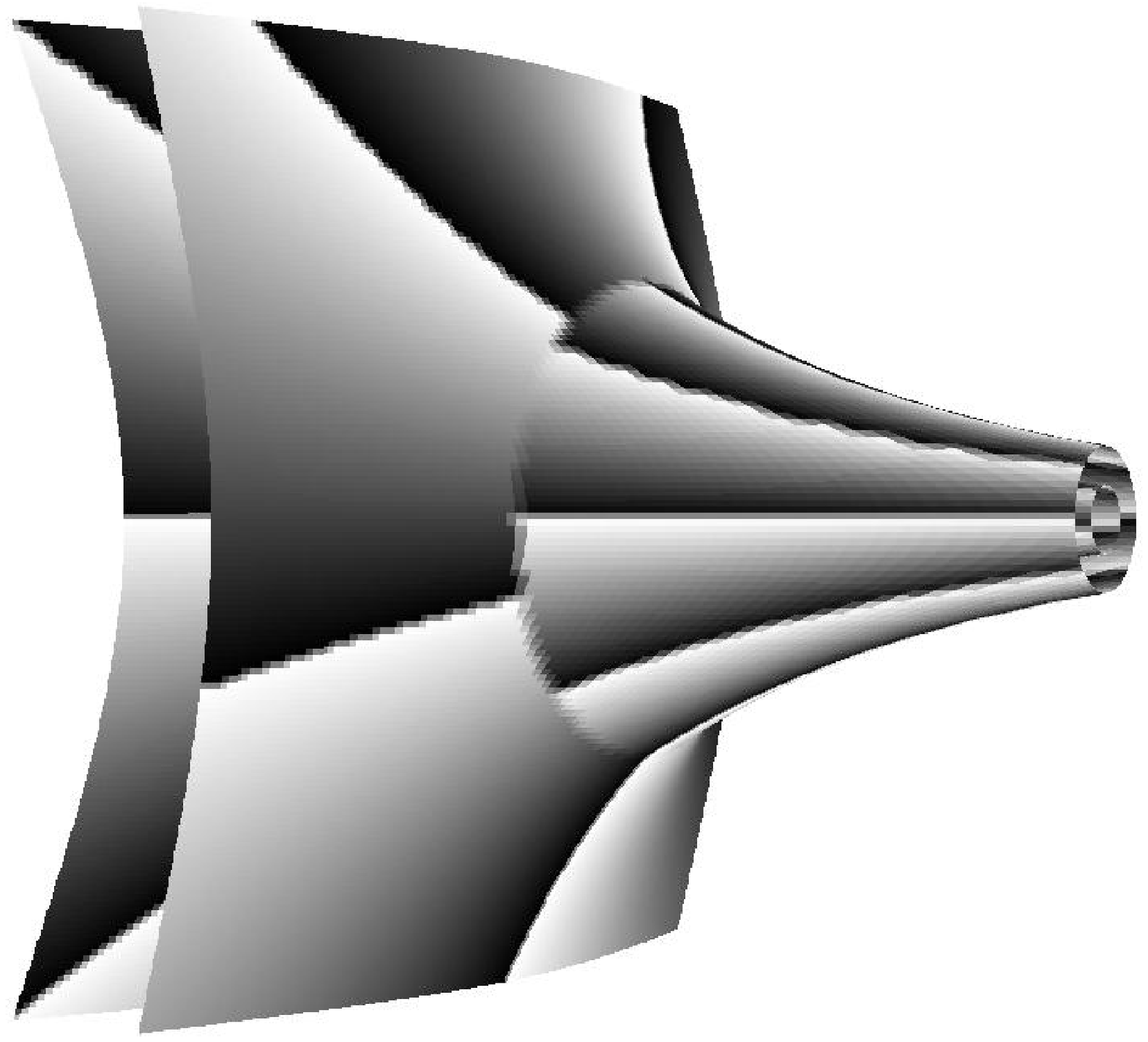}
\caption{Baryonic root of SU(8) with 13 flavors} \label{mduality}
\end{figure}

We propose a new description of Seiberg duality which has no moving
branes and is valid at any fixed, finite value of $\Lambda$.  Consider
the case $N_f > N_c$ at the root of the baryonic branch in an
$\cal{N}=$2 theory with no bare quark masses.  Then according to
Ref.~{\cite{HoriOz}} the connected components of the M5 brane are
described by
\begin{equation}
t=v^{N_c}\hsp{.5}\textup{and}\hsp{.5} t=v^{\tilde{N}_c}, \hsp{.3}
\tilde{N}_c=N_f-N_c
\end{equation}
as seen in Figs.~\ref{duality} and \ref{mduality}.  These two M5
branes intersect at $N_c-\tilde{N}_c$ points.

The crucial realization is that the reduction of this configuration to
IIA and in particular the $x^6$ location of the NS5 branes is not
uniquely defined \cite{Witten}.  Rather, we claim, that the effective
location of the NS5 branes depends on the energy scale probed, $E$.
We consider charged hypermultiplet matter corresponding to M2 branes of disk
topology stretched between the two branches of the M5 brane. It is
clear from Fig.~\ref{mduality} that the area of such an M2 brane (and
therefore its energy) is proportional to the distance in $v$ between
the two branches of M5 brane at its position in $x^6$. This restricts
the regions on the M5 brane that a quark of a given energy can
probe. Thus, for such a probe, one may effectively place an NS5 brane
at the intersection of the corresponding M5 brane and $v \sim E$. A
high energy $E_1 \gg \Lambda$ (semiclassical) M2 brane probe is
restricted to probe large features at small $x^6$. Thus, to such a
probe, the configuration consists of an NS5 brane with an NS$5_\theta$
brane on its left, $N_c$ color branes connecting them and $N_f$
semi-infinite flavor branes which extend to the right. That is, the
reduction to type IIA is Fig.~{\ref{sduality}}a, which can roughly be
obtained by drawing an NS5 brane wherever the line $E_1$ crosses an
$M5$ brane.  This corresponds to the SU($N_c$) asymptotically free
electric theory.  However, low energy quarks can only exist at
sufficiently large $x^6$.  Thus, a low energy $E_2 \ll \Lambda$ M2
brane probe corresponding to a charged quark is only sensitive to the
configuration of the M5 brane at large $x^6$, the right side of
Fig.~\ref{duality}. Thus, if we consider the portion of the M5 brane
configuration accessible to such a probe, we would find that the
corresponding reduction to IIA at $E_2$ is Fig.~{\ref{sduality}}d.

Now, note that the two M5 branes cross at the QCD scale
$\Lambda$, and at energy scales below this, that is, further right in
the figure, the M5 brane is to the left of the M5$_\theta$ brane.
Thus a probe at energies below the QCD scale will see the two M5
branes interchanged, which is the usual description of the magnetic
theory.  This theory is IR free because the branes separate as $v$
increases, and has a Landau pole where the branes cross.

\section{Flavor Symmetry Breaking} \label{flavbreak}

\subsection{Flavored Magnetic Monopoles}

Recall that a flavorless magnetic monopole (or dyon) in a IIA
realization of the electric picture is a D2 brane with disk topology
bounded by a circle which extends along one color brane, down an NS5
brane, back along another color brane and then finally back along the
other NS5 brane, as in Fig.~\ref{basic}.  More generally magnetic
monopoles may be charged under the U($N_f$) flavor symmetry.  
In this case the monopole may include fundamental strings extending from
the D2 brane to a flavor brane.  Another version of the s-rule states
that at most one such fundamental string can extend from a given
monopole to a given flavor brane, which identifies these strings as
excitations of the fermionic zeromodes present in flavored monopoles.

\begin{figure}[htb]
\centering
\includegraphics[width=6in]{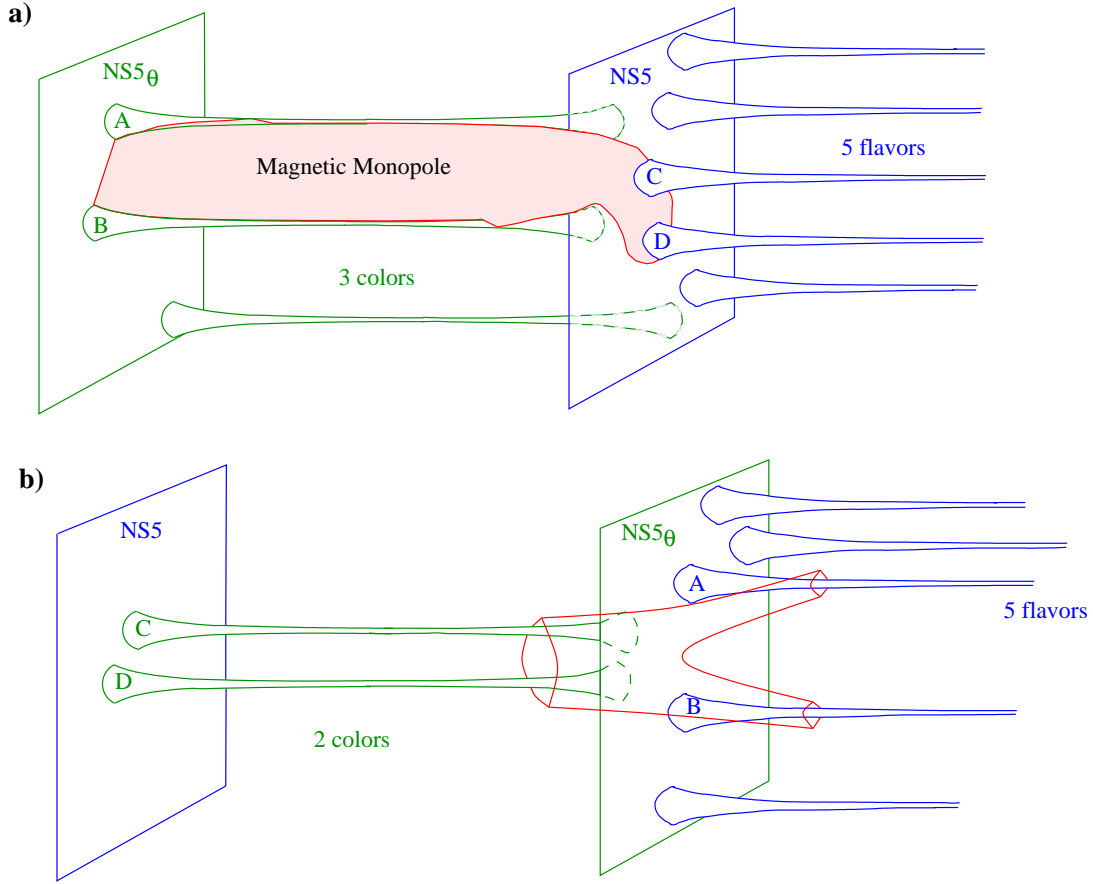}
\caption{a) A magnetic monopole in an antisymmetric tensor flavor
  representation. \hsp{.5} b) In the dual magnetic picture it becomes
  a baryon.} \label{magmon}
\end{figure}

The M theory lift of this monopole configuration is a single M2 brane
with the topology of a disk. Thus, its boundary is a circle which
wraps the M theory direction a total of zero times, as seen in
Fig.~{\ref{magmon}}.

The microscopic mechanism behind the flavor symmetry breaking of
Refs.~{\cite{Murayama1, Murayama, Murayamatalk}} is the condensation
of magnetic monopoles in an antisymmetric tensor flavor representation
drawn in Fig.~\ref{magmon}a.  The transformation properties under
U($N_f$) can be read off from the brane cartoon.  If the monopole does
not wrap any flavor branes it transforms as a flavor singlet.  A
monopole that wraps one of the flavor branes transforms in the
fundamental representation, while wrappings of more than one flavor
brane transform in the antisymmetric representation of the flavor
group. To see why this representation is antisymmetric, notice that in
IIA, if the D6 branes are moved between the NS5 branes, a monopole
consists of a D2 brane connected by strings to D6 branes.  The s-rule
provides an exclusion principle, restricting the number of strings
connecting a monopole to a D6 brane to 0 or 1. As a consistency check
on this picture, notice that there are $2^{N_f}$ configurations of
wrappings which agrees with the known number of states in the
representation.

In order to understand how semiclassical magnetic monopoles in the UV
theory are related to the IR degrees of freedom, we can consider an M2
brane corresponding to a high energy monopole configuration and follow
its decay into the IR. Deforming the UV theory away from the baryonic
root to clearly visualize its charges, such a monopole is pictured in
Fig.~\ref{magmon}a. We deform back to the baryonic root and then allow
it to decay, requiring that its wrappings (i.e. charges) are
preserved. Its energy will become sufficiently low that it is best
described using the dual magnetic description, that is, with the NS5
branes switched. Now, if we deform the IR theory to a generic point in
its Coulomb branch and keep track of the wrappings, this configuration
corresponds to magnetic baryons as seen in Fig.~\ref{magmon}b.  The
magnetic theory is IR free, and so these baryons decay into magnetic
quarks whose condensation provides the order parameter of the flavor
symmetry breaking.

It is also easy to see that the correlation between the electric
charge and chirality of dyons discussed for SU(2) gauge theories in
Ref.{\cite{SeibergWitten2}} follows from the fact that the monopole is
topologically a disk.  All monopoles with even chirality come from
monopoles whose boundary wraps an even number of flavor branes and
therefore the M theory direction an even number of times along the M5
brane as well.  Each flavor brane wrapping introduces a
hypermultiplet into the antisymmetric representation and hence the
monopole has chirality $(-1)^H=1$.  Each color brane wrapped yields a
unit (using the conventions of {\cite{SeibergWitten2}}) of electric
charge and so the monopole acquires an even electric charge.  This
argument works similarly for odd chirality and odd electric charge.

\subsection{Symmetry Breaking Pattern}

As we reviewed in section \ref{sumfield}, field theory
calculations in a variety of limits showed that flavor symmetry is
generically broken
\begin{equation}
U(N_f)\hsp{.1}\rightarrow\hsp{.1}U(r)\times U(N_f-r)
\end{equation}
in softly broken $\N=1$ asymptotically free ($N_f<2N_c$) SQCD in the
limit that bare quark masses vanish.  We will draw string and M theory
realizations of these limits and use them to reproduce the qualitative
results of several field theory calculations.  In particular, in the
semiclassical limit, we will relate $r$ to the number of color branes
attached to flavor branes.

\begin{figure}[htb]
\centering
\includegraphics[width=6in]{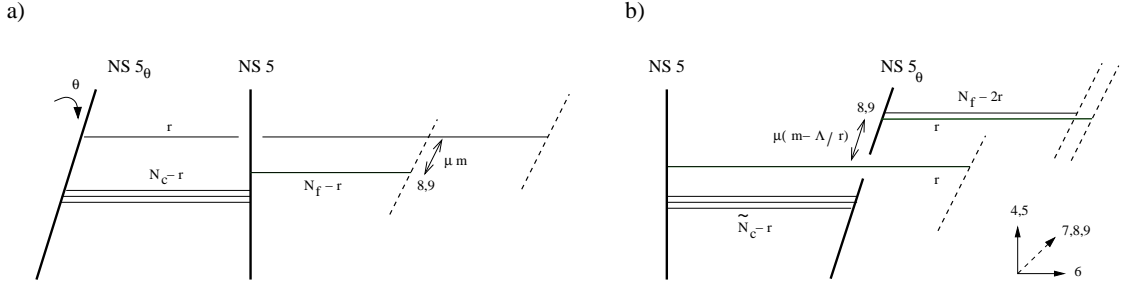}
\caption{a) Semiclassically when SUSY is broken by rotating the
  NS$5_\theta$ brane, $r$ color branes connect to flavor branes while
  the rest slide to $v=0$.\hsp{.4}b)The dual magnetic description of
  the nonbaryonic branches can be understood similarly to the
  semiclassical case.} \label{semivevs}
\end{figure}

\subsection{Semiclassical Analysis}

Following Refs.~{\cite{Murayama1, Murayama, Murayamatalk}} we begin by
considering bare quark masses much larger than the QCD scale so that a
semiclassical (IIA) analysis is valid.  This will allow us to count
the total number of vacua for comparison with later calculations.
Recall from section~\ref{N1section} that after rotating the
NS$5_\theta$ brane all color branes must either slide to $v=0$ or
connect to a flavor brane as in Fig.~\ref{semivevs}a.  The number of
color branes connecting to flavor branes, $r$, clearly can neither
exceed the number of color branes nor the number of flavor branes. We
illustrate the simple case of the vacua arising this way with the bare
quark masses all equal $m=m_i \gg \Lambda$ in Appendix A
for the case of SU(3) with $N_f=4$.


To count the number of vacua with generic quark masses, notice that
$r$ flavor branes can attach to color branes in $(^{N_f}_{\ r})$ ways
(recall that there is no combinatoric factor from choosing which color
branes to attach as these choices are gauge equivalent), leaving
$N_c-r$ color branes which form a line\footnote{Actually they form an
  ellipse whose semi-minor axis scales with $\theta$.  This ellipse
  degenerates to a line when $\theta=0$ and a circle with an
  $A_{N_c-r-1}$ singularity in its center at $\theta=\pi/2$.}
{\cite{HananyZaffaroni}} centered at $v=0$, as shown in
Fig.~\ref{Uday}.  This line can be rotated by integer multiples of
$\pi/(N_c-r)$ without affecting the $x^{10}$ coordinates
asymptotically far away, corresponding to the anomaly-free R-symmetry
subgroup.  The $N_c-r$ inequivalent orientations of the line result in
$N_c-r$ different vacua, in agreement with the Witten index of this
theory.  Thus the total number of semiclassical vacua (assuming all
bare quark masses are distinct and nonvanishing) is
\begin{equation}
\N_{sc}=\sum_{r=0}^{min(N_c+1,N_f)}(N_c-r)(^{N_f}_{\ r})
\end{equation}
in agreement with computations from field theory considerations in
Refs.~{\cite{Murayama1, Murayama, Murayamatalk}}.

\subsection{Nonbaryonic Branches}

Semiclassically flavor symmetry is broken by meson vevs equal to $\mu
m_i \sim \mu \ \textup{tan}\theta$ which is the distance in the $w$
plane shown in Fig.~\ref{semivevs}a.  These vevs vanish when the bare
quark masses vanish, apparently restoring the explicitly broken flavor
symmetry.  However we will see that, if we include quantum effects, in
some vacua the flavor symmetry remains broken even in this limit.  For
simplicity let all of the bare quark masses be equal $m=m_i \ll
\Lambda$.  

\begin{figure}[htb]
\centering
\includegraphics[width=4in]{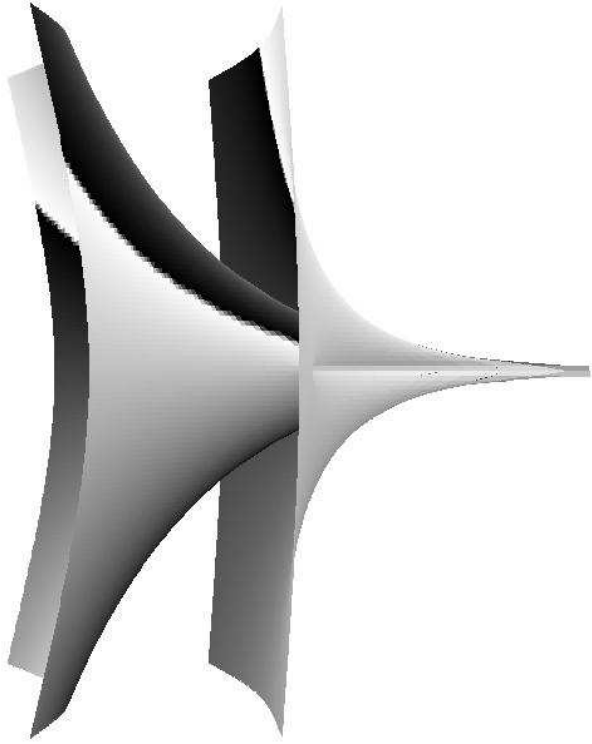}
\caption{This M5 brane configuration corresponds to the $r=2$
  nonbaryonic root of SU(3) with 4 flavors.  In the IR (the right side
  of the picture) a reduction to IIA produces two parallel, almost
  coincident NS5 branes indicating that the theory is strongly
  coupled.  The fact that the distance between the branes converges
  indicates that the IR theory is superconformal.}  \label{scir}
\end{figure}

\begin{figure}[htb]
\begin{minipage}[c]{.25\textwidth}
\centering
  \includegraphics[width=1.6in]{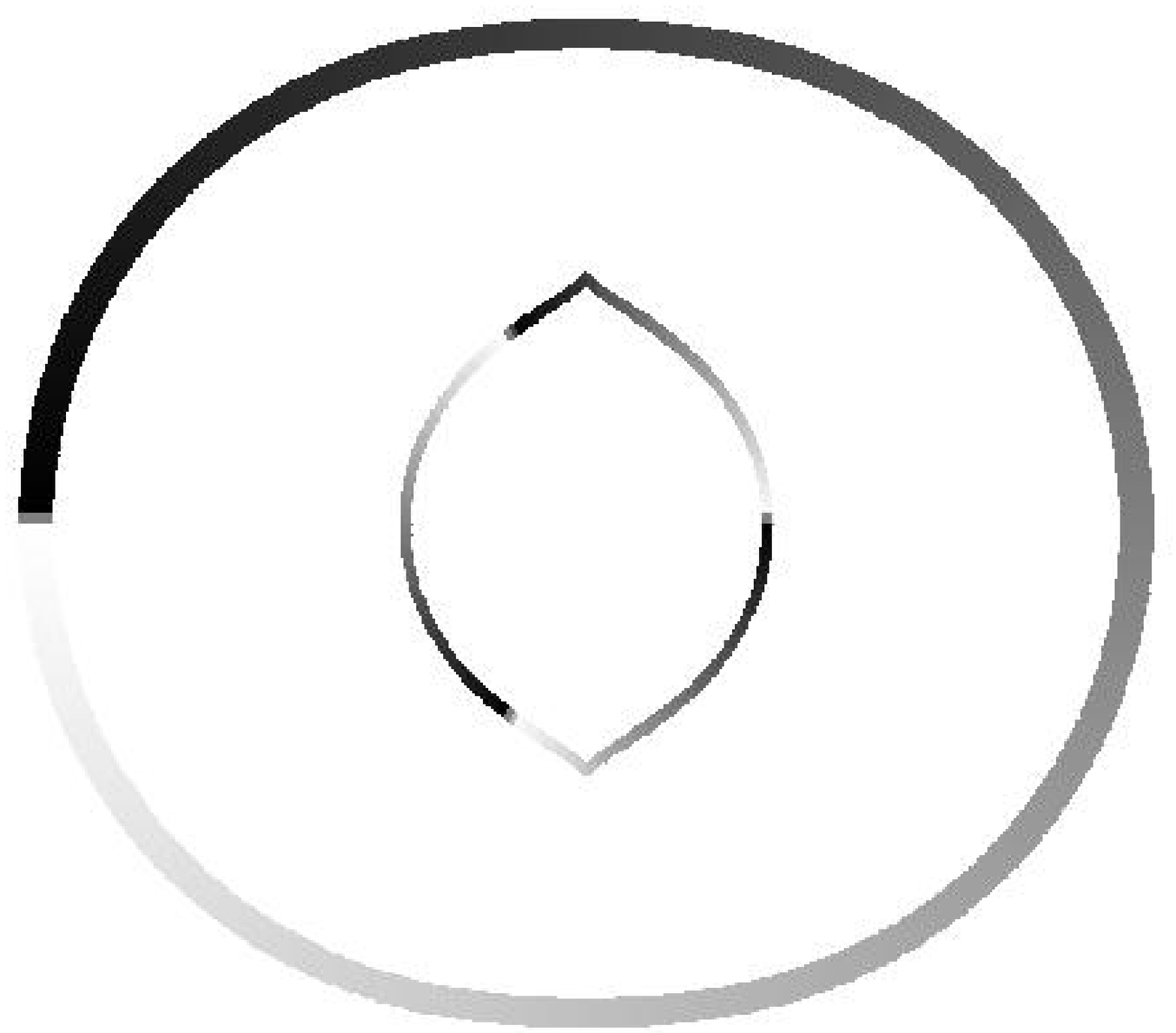}
\end{minipage}%
\begin{minipage}[c]{.25\textwidth}
\centering
  \includegraphics[width=1.6in]{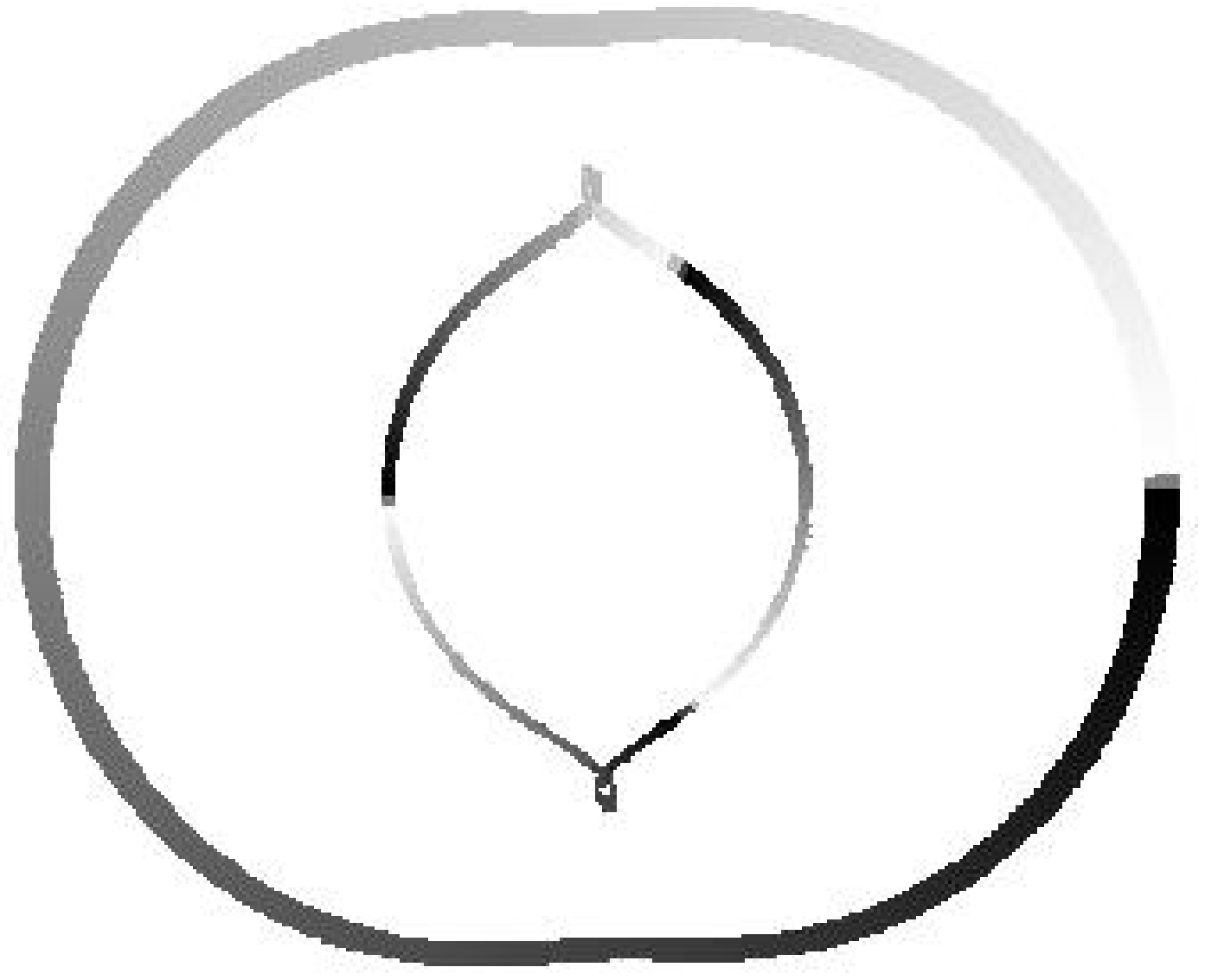}
\end{minipage}%
\begin{minipage}[c]{.25\textwidth}
\centering
  \includegraphics[width=1.6in]{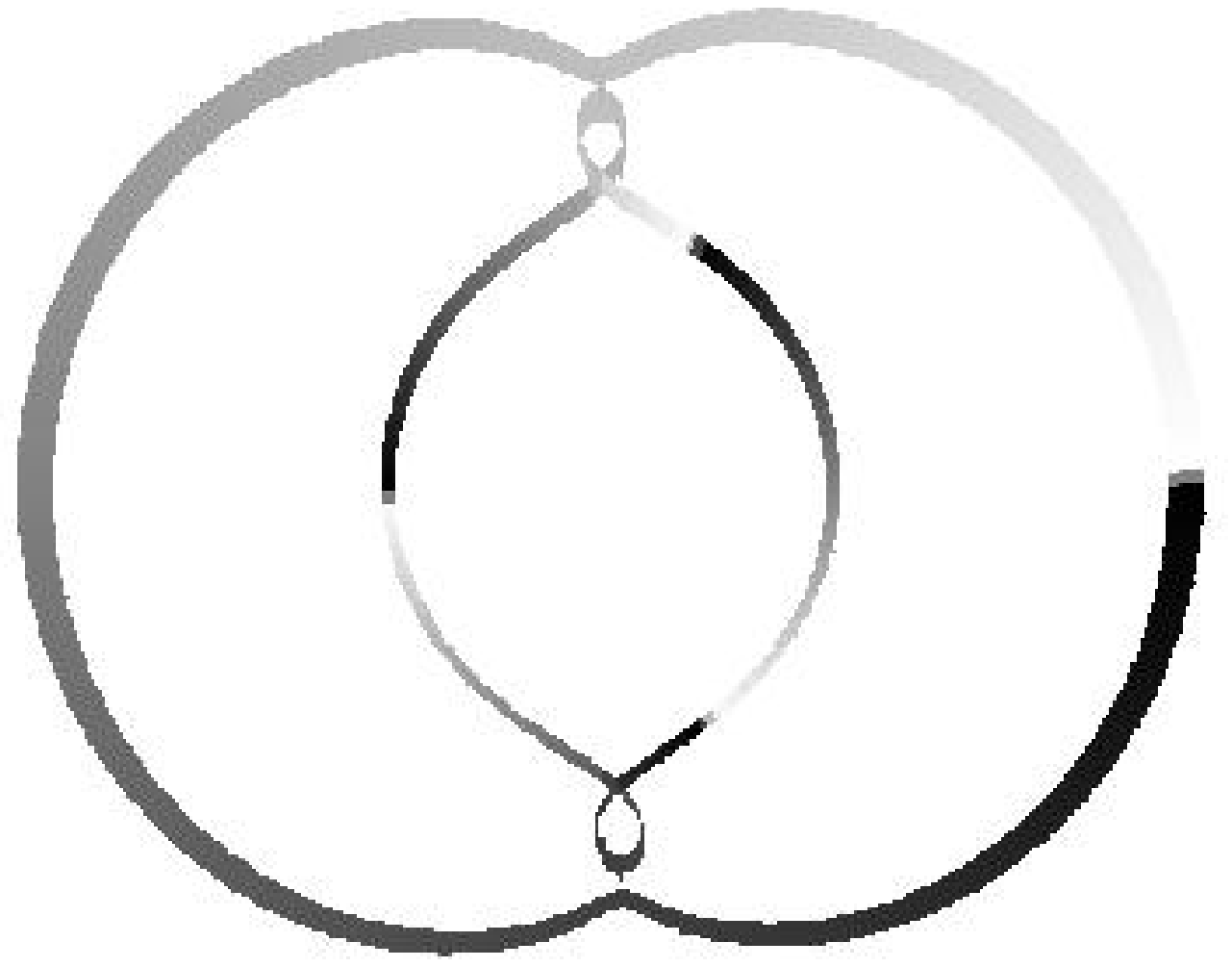}
\end{minipage}%
\begin{minipage}[c]{.25\textwidth}
\centering
  \includegraphics[width=1.6in]{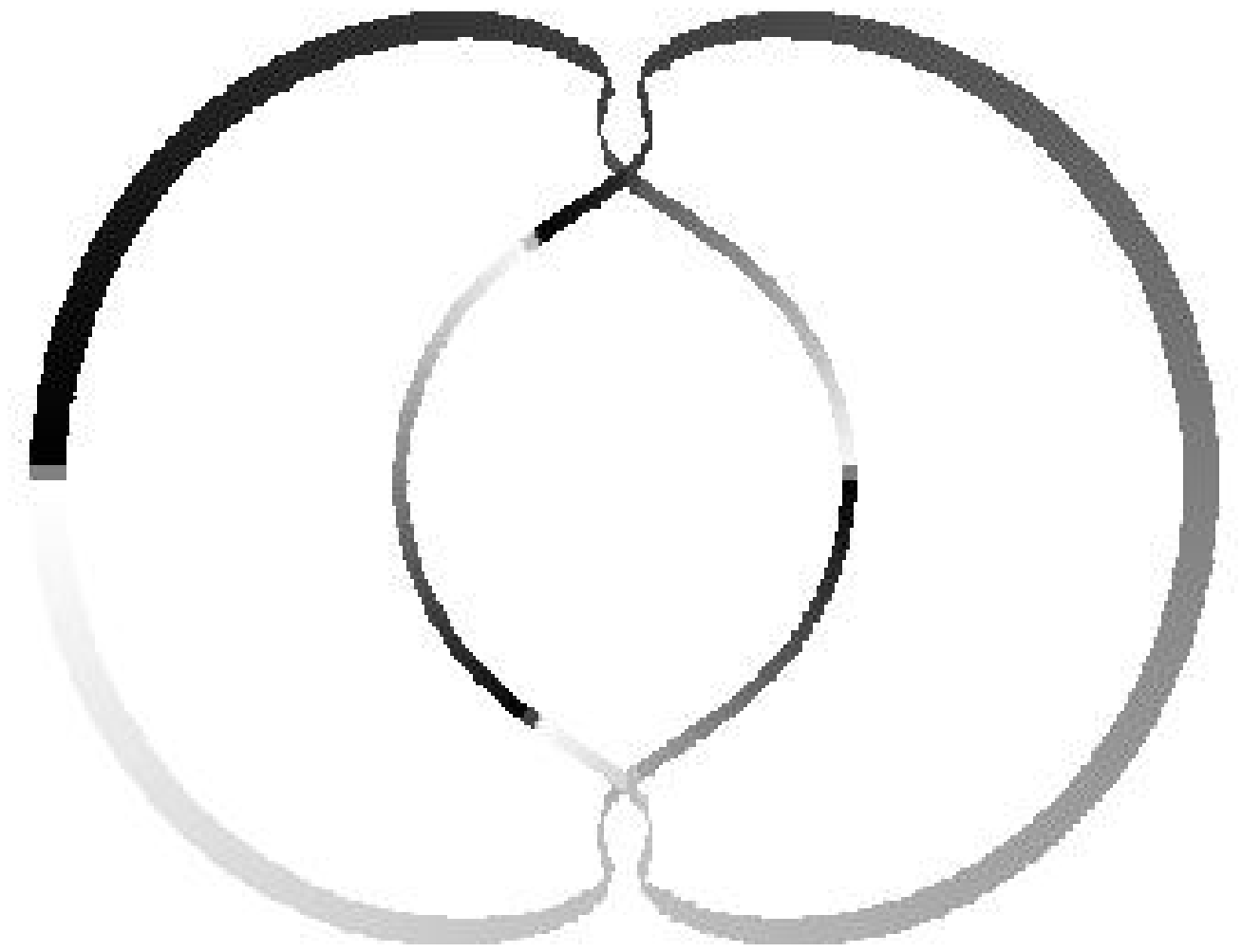}
\end{minipage}
\begin{minipage}[c]{.25\textwidth}
\centering
  \includegraphics[width=1.2in]{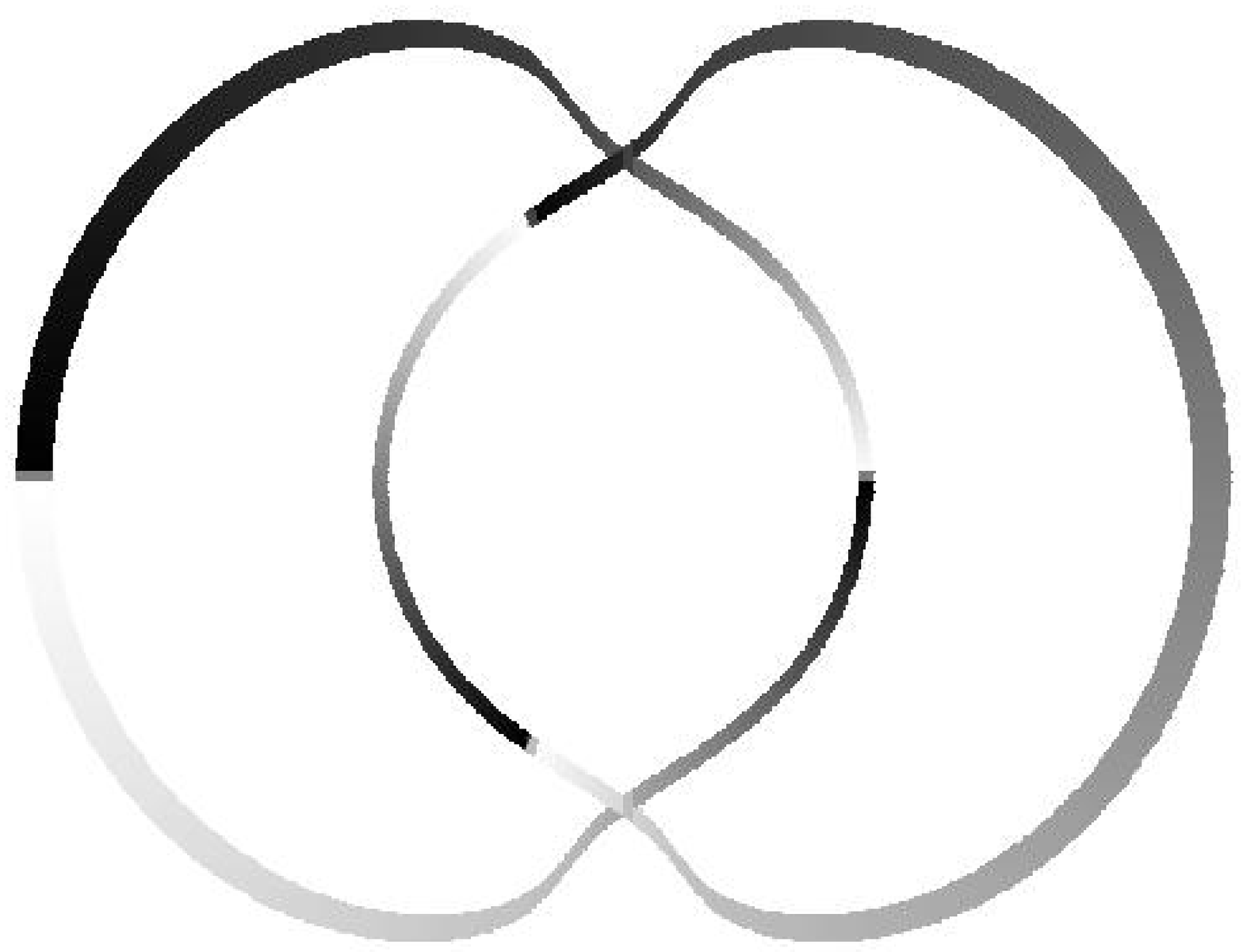}
\end{minipage}%
\begin{minipage}[c]{.25\textwidth}
\centering
  \includegraphics[width=1.5in]{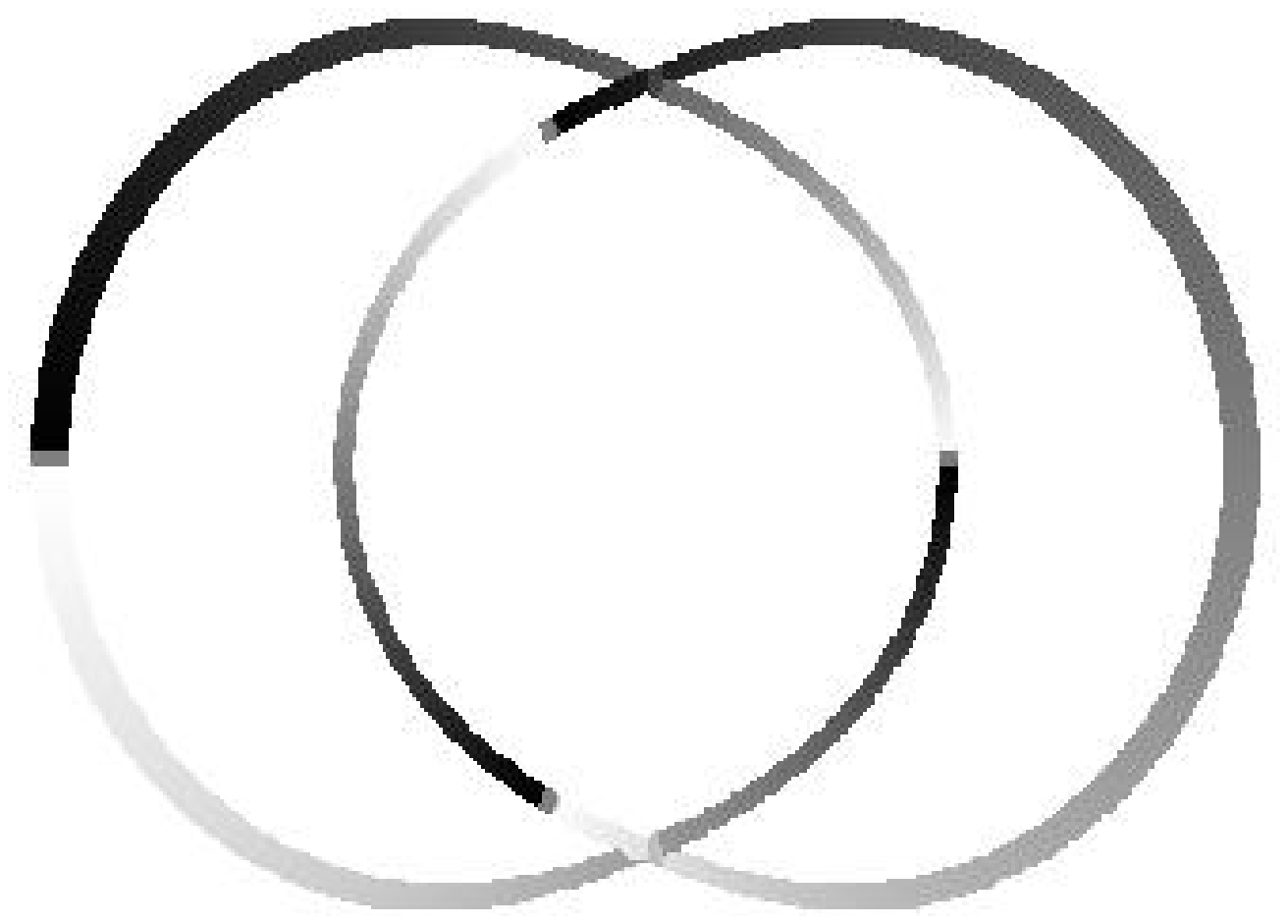}
\end{minipage}%
\begin{minipage}[c]{.25\textwidth}
\centering
  \includegraphics[width=1.5in]{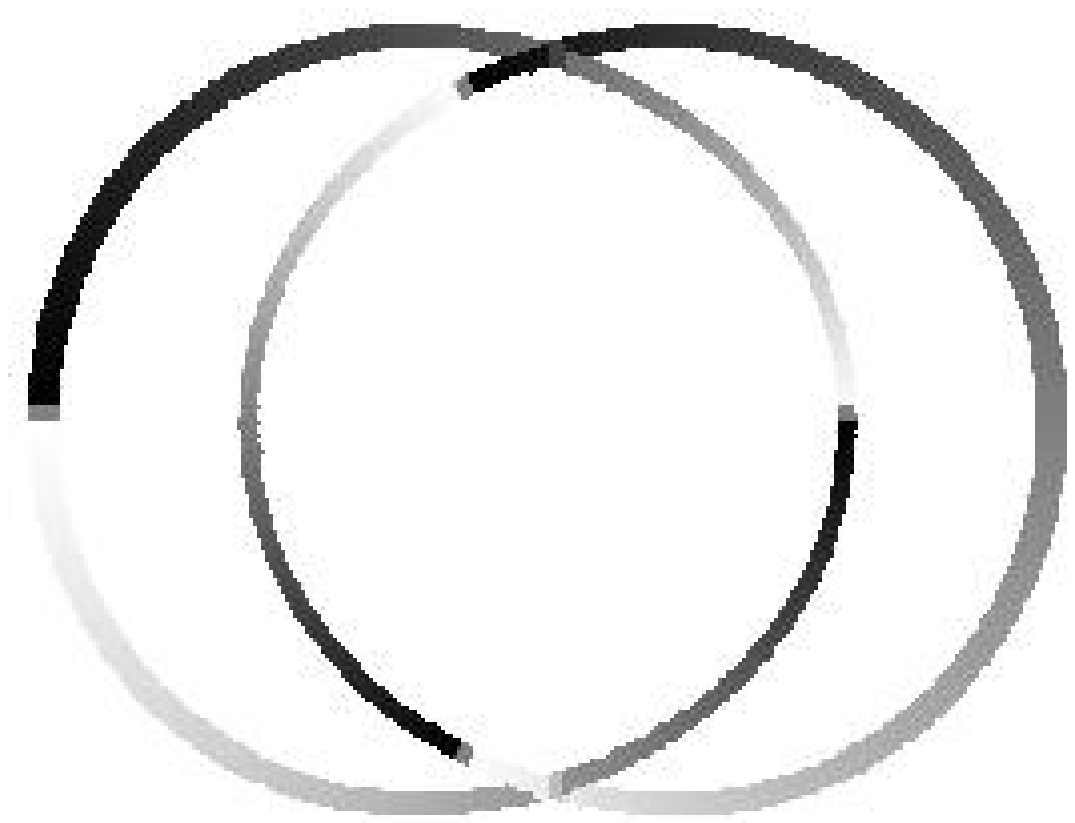}
\end{minipage}%
\begin{minipage}[c]{.25\textwidth}
\centering
  \includegraphics[width=1.2in]{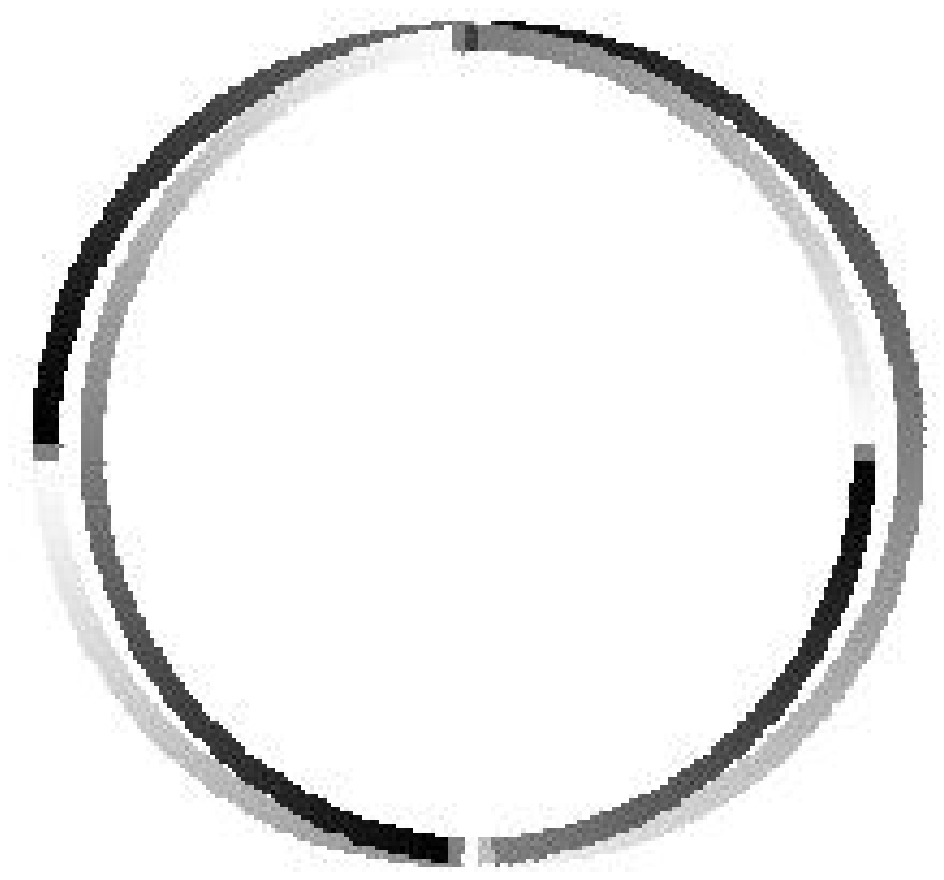}
\end{minipage}
\caption{Above are M5 brane cross-sections at constant $x^6$
  corresponding roughly to constant energy scales of the $r=2$
  nonbaryonic root of the SU(3) theory with 4 flavors and no bare
  quark masses.  Reading from left to right and top to bottom the
  energy scale of the corresponding M2 brane probes decreases.  The M
  theory direction is parameterized by darkness.}  \label{scircs}
\end{figure}

The $r=N_f/2$ theory is superconformal in the IR and generally
difficult to understand. An example of M5 brane associated with such a
vacuum is drawn in Fig.~\ref{scir}.  The two branches, corresponding
to different NS5 branes upon reduction to IIA, intersect at exactly
two points.  Cross-sections to the left and right of these two
singularities are seen in the first and second lines of
Fig.~\ref{scircs}.  They are topologically distinct and correspond to
distinct reductions to IIA, yielding different effective field
theories.

We will be interested in the IR free case of $0<r<N_f/2$,
corresponding to nonbaryonic branches.  The fundamental degrees of
freedom must be magnetic because in the UV this theory has a Landau
pole (M5 branes cross) separating it from the semiclassical region
with electric degrees of freedom.  Thus flavor symmetry breaking can
only be caused by magnetic quark vevs, which like meson vevs (meson
vevs are quadratic in quark vevs) correspond to distances in the $w$
plane.  Semiclassically this distance was $\mu m_i$ and so vanished
when the bare quark masses were taken to zero.  Quantum mechanically
these flavor branes have width $O(\Lambda)$, critically changing the
distances between them.  As a result the $r$ nonvanishing magnetic
quark vevs
\begin{equation}
q=\sqrt{\mu(m_i-\Lambda/r)}
\end{equation}
do not vanish when the bare masses are eliminated, as seen in
Fig.~\ref{semivevs}b.  These quark vevs break the flavor symmetry
\begin{equation}
U(N_f)\ \rightarrow\ U(r)\times U(N_f-r).
\end{equation}
These vevs also break the global $\Z_{2N_c-N_f}$ symmetry and
so there must be $2N_c-N_f$ copies of this configuration.  When $r=0$,
$q=\sqrt{\mu m_i}$ and flavor symmetry is unbroken.  Again there are
$(^{N_f}_{\ r})$ ways to choose $r$ quarks, leading to a total of
\begin{equation}
\N_1=(N_c-\tilde{N}_c)2^{N_f-1}
\end{equation}
vacua of this type.  Notice that when $N_f<N_c$, $\N_1$ agrees with
$\N_{sc}$ and therefore by supersymmetry this is a complete
classification of the vacua.

\subsection{Baryonic Branch}
\begin{figure}[htb]
\centering
\includegraphics[width=6in]{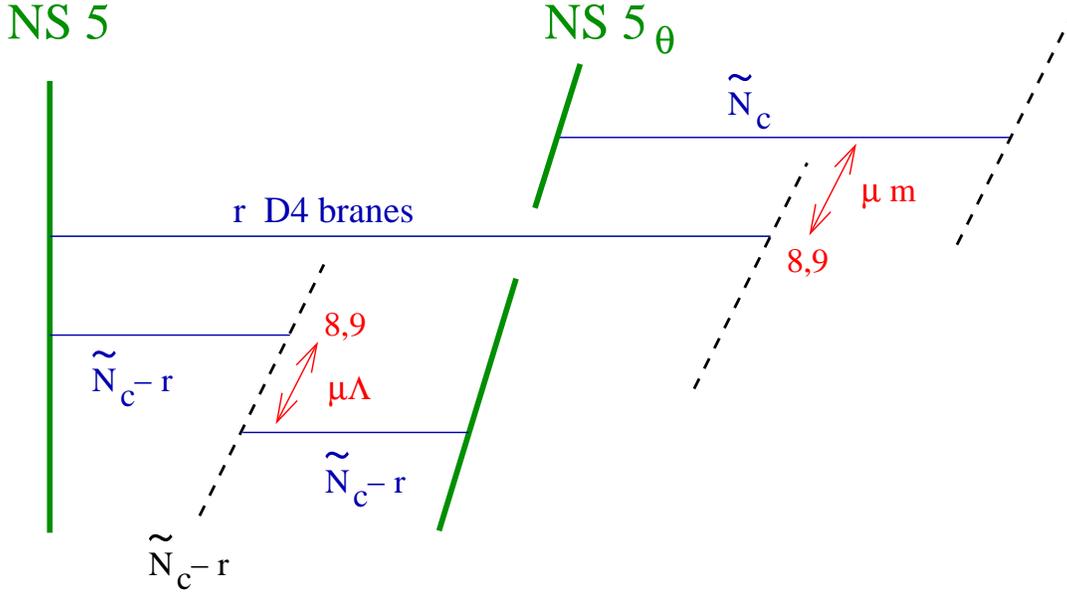}
\caption{On the baryonic branch, the baryon vev $\mu\Lambda$ is
  realized as a relative slide of the two halves of the configuration
  along a collection of D6 branes.} \label{brootfig}
\end{figure}

The above analysis is incomplete when $N_f\ge N_c$ because every color
brane can be broken by a D6 brane and the two halves displaced from
each other along the $w$ plane by a distance (baryon vev) exactly
canceling the displacement measured by the dual quark vev.  Clearly
this requires $N_f\ge N_c$ because $N_f$ is the number of D6 branes
while $N_c$ is the number of color branes, and each color brane
requires a D6 brane along which to break, as illustrated in
Fig.~\ref{brootfig}.

More concretely, at the root of the baryonic branch the unattached
$N_c-\tilde{N}_c$ color branes form a circle.  This means that
$N_c-\tilde{N}_c$ magnetic monopoles (or dyons), which are between
adjacent color branes, become massless and can acquire vevs.  The
baryonic branch has one more massless monopole compared to the
nonbaryonic branch, which is enough to completely Higgs the
U(1$)^{N_c-\tilde{N}_c}$, and therefore there are enough vevs to
control the center of mass motion of this set of color branes.  The
center of mass mode of the entire system is infinitely massive and so
a shift in the center of mass of the $N_c-\tilde{N}_c$ branes leads to
an opposite center of mass shift of the Seiberg dual SU($\tilde{N}_c$)
gauge theory along the $w$ plane, the same shift parametrized by all
quark vevs.  One result is that the magnetic monopoles are charged
under the U(1$)_B$ baryon number.  The crucial implication is that a
slide along the $w$ plane in the U(1$)^{N_c-\tilde{N}_c}$ system can
undo the $w$-shift in the SU($\tilde{N}_c$) configuration which led to
the flavor symmetry breaking quark vev.  Thus flavor symmetry is
unbroken on the baryonic branch.

These geometrical quantities can be related to the corresponding field
theory calculation ~\cite{Murayama1, Murayama, Murayamatalk} by
considering the following superpotential terms {\cite{Plesser}} in the
magnetic description of the baryonic root
\begin{equation}
W\supset
\frac{1}{\tilde{N}_c}\textup{Tr}(q\tilde{q})(\sum_{k=1}^{N_c-\tilde{N}_c}\psi_k)-\sum_{k=1}^{N_c-\tilde{N}_c}\psi_k
e_k\tilde{e}_k+\mu\Lambda\sum_{k=1}^{N_c-\tilde{N}_c}x_k\psi_k.
\end{equation}
Here $q$ and $\tilde{q}$ are magnetic quarks, $e_k$ and $\tilde{e}_k$
are flavorless magnetic monopoles, $x_k$ are constants and $\psi_k$
are the dual photons of the abelian part of the
SU($\tilde{N}_c$) $\times$ U(1$)^{N_c-\tilde{N}_c}$ dual gauge group.
From the superpotential we see that each magnetic monopole is charged
under a U(1) while the dual quarks are charged under all of the
U(1)'s.\footnote{This is in contrast to the nonbaryonic root, where
  there is one less massless magnetic monopole and so after a basis
  change the dual quarks and magnetic monopoles are charged under
  disjoint U(1)'s.}  As a result the D term equation for each U(1) is
\begin{equation}
\frac{1}{\tilde{N}_c}\textup{Tr}(q\tilde{q})-e_k\tilde{e}_k+\mu\Lambda
x_k = 0. \label{dequation}
\end{equation}
This means that, because there are as many $e_k$'s as $x_k$'s,
Tr$(q\tilde{q})$ can vanish if each $e_k\tilde{e}_k$ is chosen
correctly and, in fact, this solution is consistent with the rest of
the D and F term constraints.  Thus we see that the D term equation
for the difference of two vevs is interpreted in M theory as the
following statement.  If the connected components of the M5 brane
slide apart in the $w$ plane along the $D6$ branes bisecting the
$\tilde{N}_c-r$ color branes then, because the components are rigid,
they also separate along the $w$ plane at the semi-infinite flavor
branes.  In other words by trading a monopole vev for a magnetic quark
vev, we find vacua which preserve flavor symmetry.  These two
distances, whose differences are preserved, are marked with
double-headed arrows in Fig.~\ref{brootfig}.

To count vacua on this branch, consider the dual SU($\tilde{N}_c$)
theory whose gauge symmetry is broken by the adjoint scalar vevs of
the $r$ color branes attached to flavor branes.  As in the
semiclassical case, the remaining $\tilde{N}_c-r$ color branes form a
line which can have $\tilde{N}_c-r$ orientations preserving the M
theory coordinate asymptotically far away, yielding a multiplicity of
$\tilde{N}_c-r$ times the combinatoric factor of $(^{N_f}_{\ r})$ from
the choice of which flavor branes to connect.  In all, this provides
\begin{equation}
\N_2=\sum_{r=0}^{\tilde{N}_c}(\tilde{N}_c-r)(^{N_f}_{\
  r})=\N_{sc}-\N_1
\end{equation}
states, and thus completes the classification of vacua.

\section{Conclusion}

We have constructed a new realization of Seiberg duality that relies
on an energy scale dependent reduction of M theory to IIA.  We have
found the M2 branes that correspond to flavored magnetic monopoles and
argued that they correspond to magnetic baryons in the dual magnetic
theory, which in turn decay to magnetic quarks.  And finally we have
interpreted baryon vevs as the relative sliding of two halves of an M
theory configuration along a Taub-NUT singularity.

As an application of the above constructions, we have reproduced the
field theory results of \cite{Murayama}.  In particular we have
correctly reproduced the flavor symmetry breaking patterns, the order
parameters of the symmetry breaking and the counting of states in
various regimes. We can interpret these countings in terms of discrete
rotations of a line of M5 brane in the $v$ plane. 

The case $r=N_f/2$ is difficult to analyze using traditional field
theory techniques, as it is superconformal and strongly coupled in the
IR.  However it is possible that by deforming the corresponding curve,
an analysis similar to the Seiberg duality of monopoles above may be
possible in this M theory setting.

A natural next step is to add orientifold planes and try to repeat
this analysis for SO($N$) and SP($N$) gauge theories.

\noindent {\bf Acknowledgements}

We would like to thank Dan Brace, Gordon Chalmers, Sergei Cherkis, Jan
de Boer, Scott Thomas, Angel Uranga, Ed Witten and Bruno Zumino for
enlightening conversations.  The work of HM and UV was supported in
part by the Director, Office of Science, Office of High Energy and
Nuclear Physics, Division of High Energy Physics of the U.S.
Department of Energy under Contract DE-AC03-76SF00098 and in part by
the National Science Foundation under grant PHY-95-14797.

\appendixa

\subsection{An Example, SU(3) with $N_f =4$} \label{Example}

We will consider in some detail the M5 branes associated with the
$r$-vacua which survive breaking to $\N=1$ for the case of SU(3) with
$N_f=4$ and equal quark masses, $m_i=m$. This case is rather
special in that the coordinates of these vacua in the moduli
space can be analytically determined as a function of mass. Thus, we
can explicitly draw out the corresponding M5 brane configurations and
test our claims for the brane interpretations of the $r$ vacua in
various regimes. 

\begin{figure}[htbp]
\centering
\begin{minipage}[c]{0.50\textwidth}
{(a) The baryonic root}
\centering
\includegraphics[width=3in]{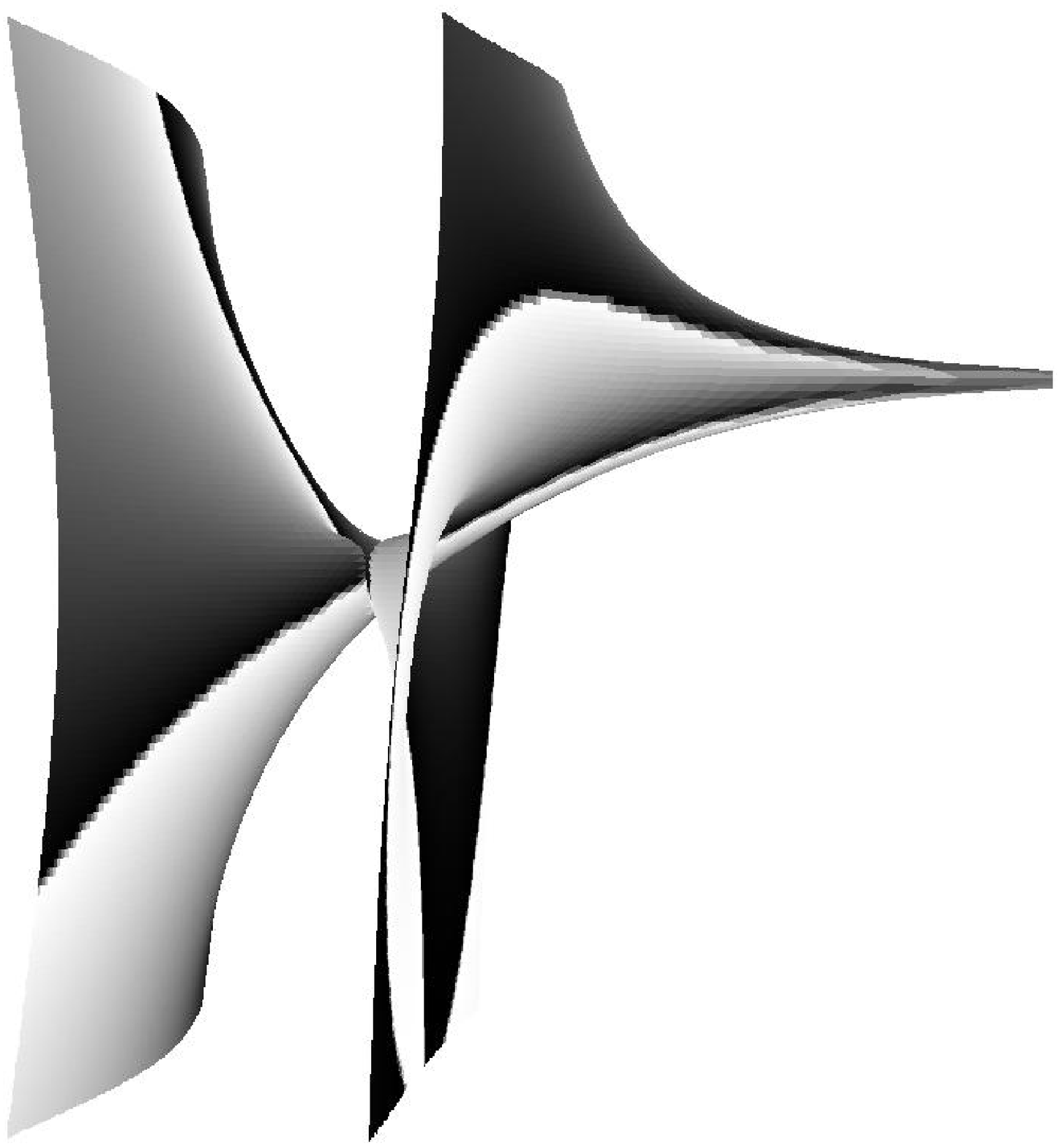}
\end{minipage}%
\begin{minipage}[c]{0.50\textwidth}
{(b) A non-baryonic r=0 root}
\centering
\includegraphics[width=3in]{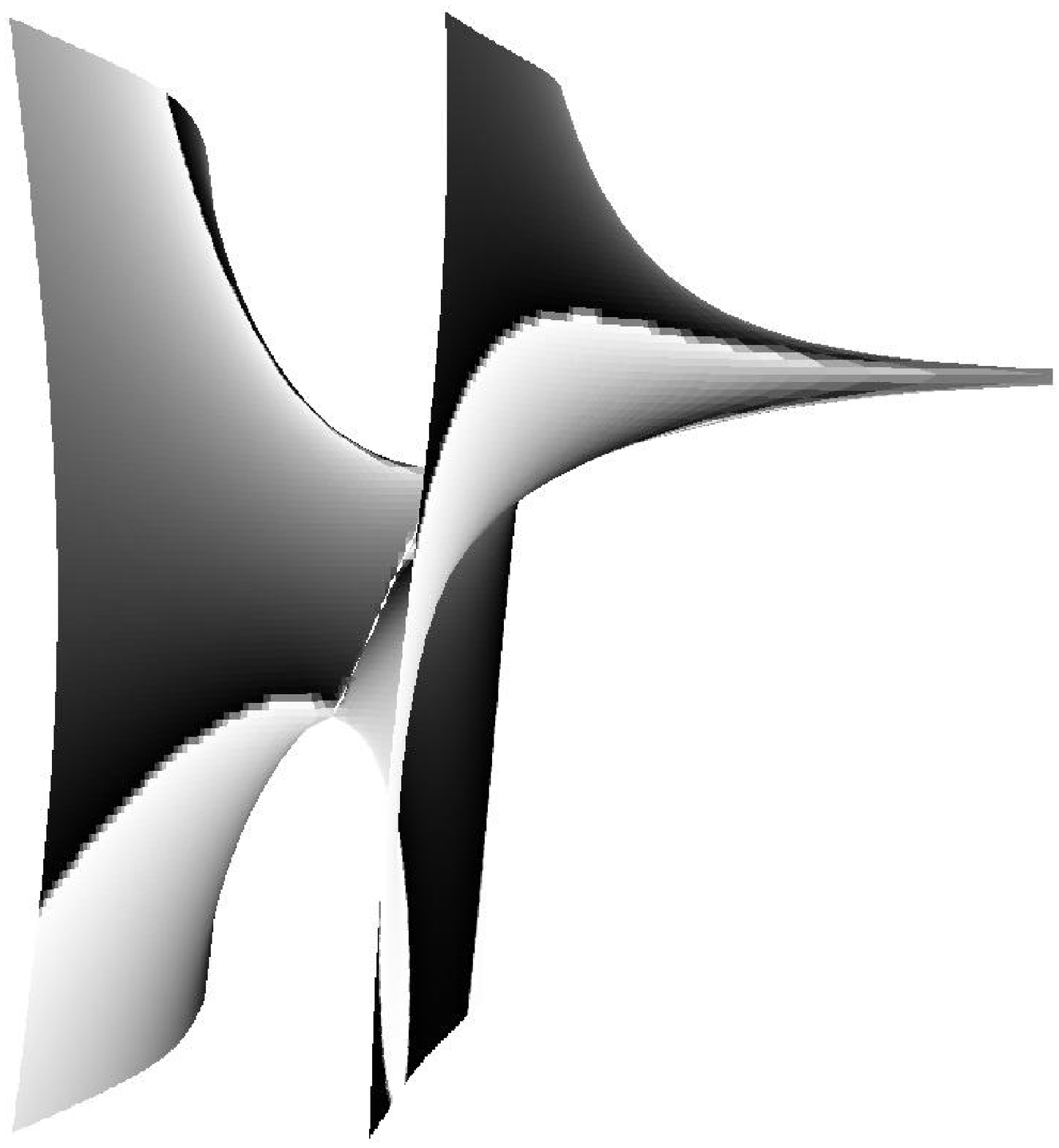}
\end{minipage}
\begin{minipage}[c]{0.50\textwidth}
{(c) A non-baryonic r=1 root}
\centering
\includegraphics[width=3in]{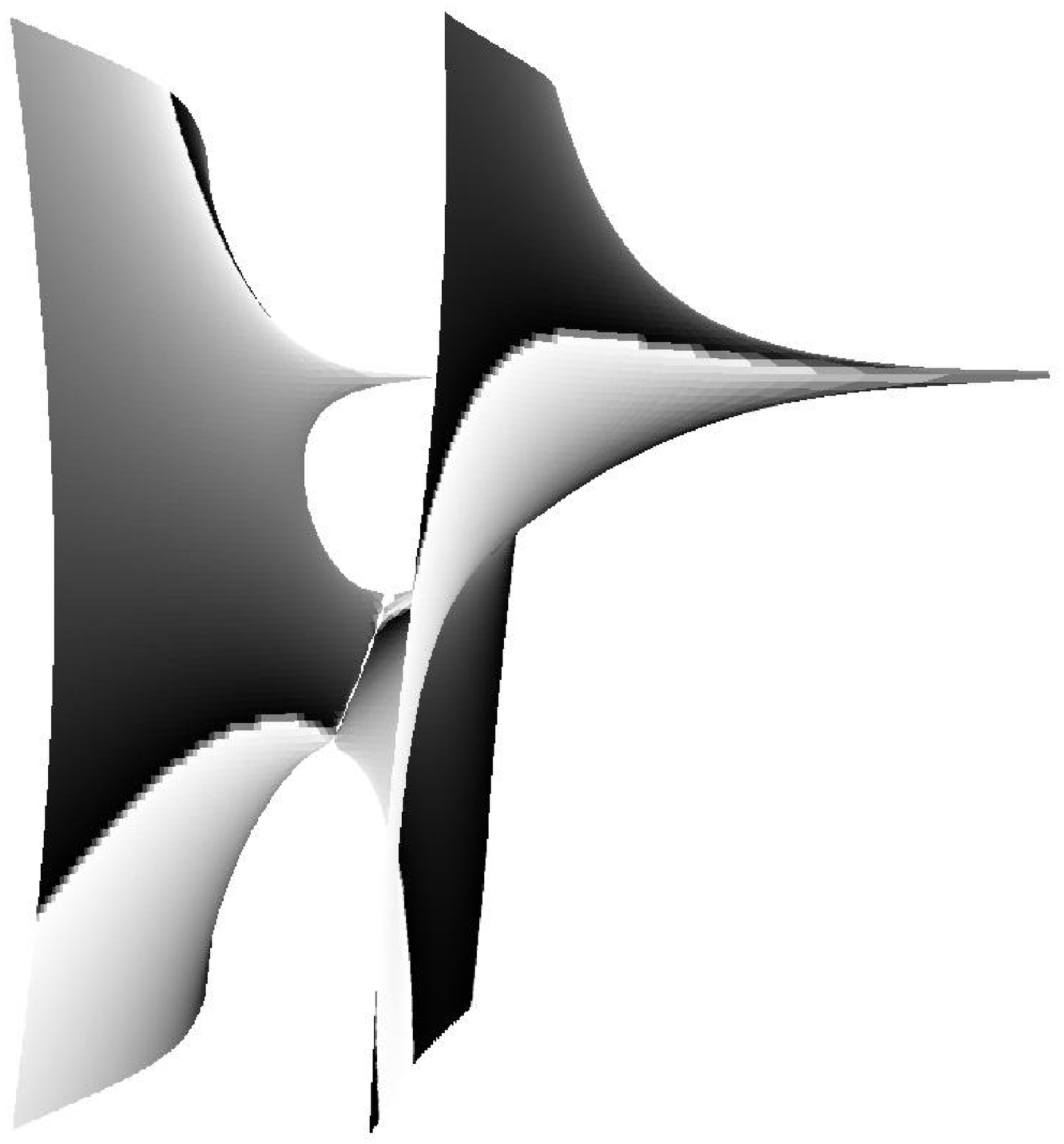}
\end{minipage}%
\begin{minipage}[c]{0.50\textwidth}
{(d) The non-baryonic r=2 root}
\centering
\includegraphics[width=3in]{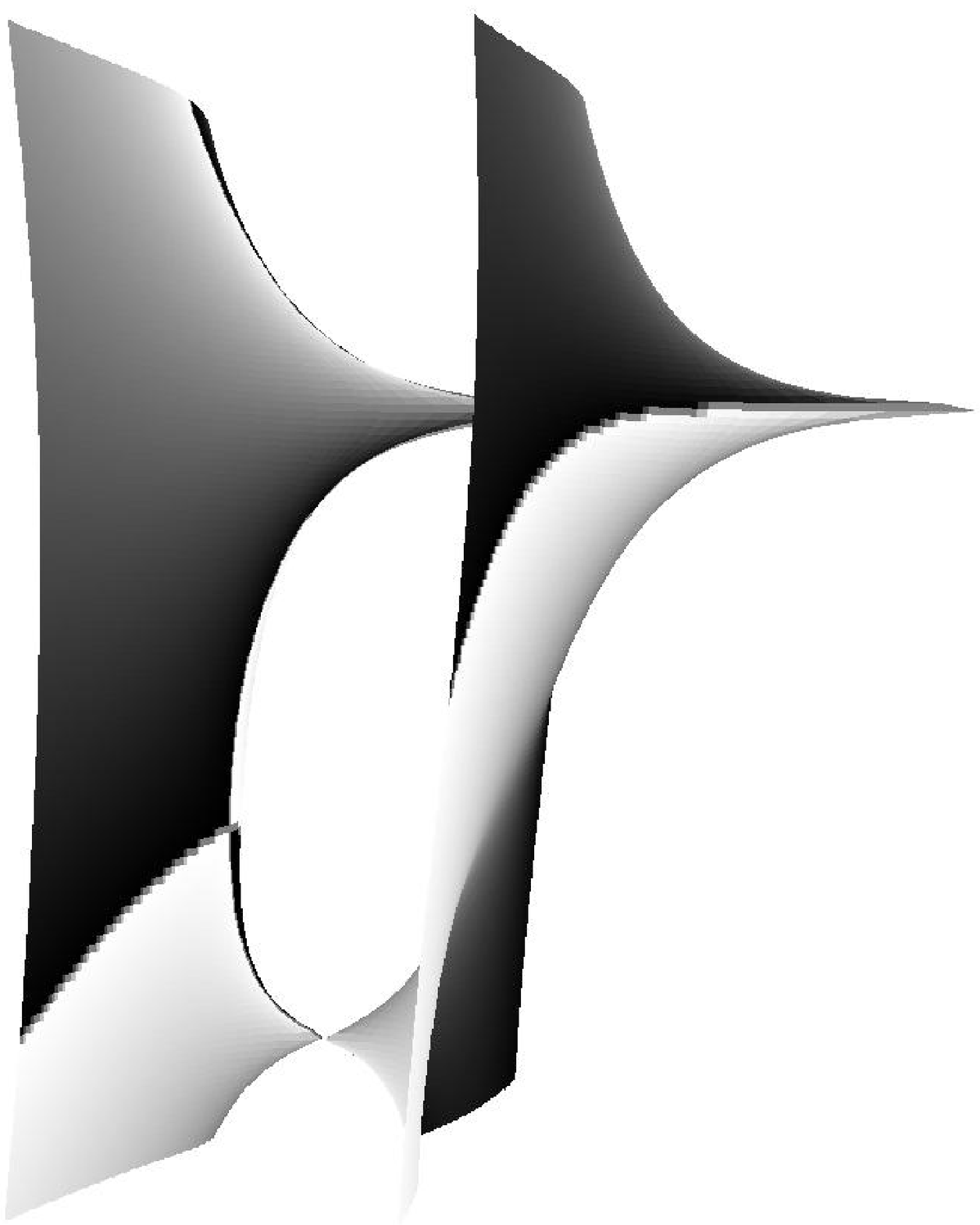}
\end{minipage}
\caption{The M5 brane configurations corresponding to $r$-vacua in SU(3)
  with 4 flavors in the semiclassical limit, i.e. with $m_i = m \gg
  \Lambda $. The flavor branes and color branes which have connected
  are represented by semi-infite tubes extending from the left branch
  of the M5 brane. The connections between the two NS5 branes are the
  remaining color branes, pairs of which have condensed massless
  monopoles between them.}  \label{semicm5}
\end{figure}

\begin{figure}[htbp]
\centering
\begin{minipage}[c]{0.333\textwidth}
{(a) The baryonic root}
\centering
\includegraphics[width=2.2in]{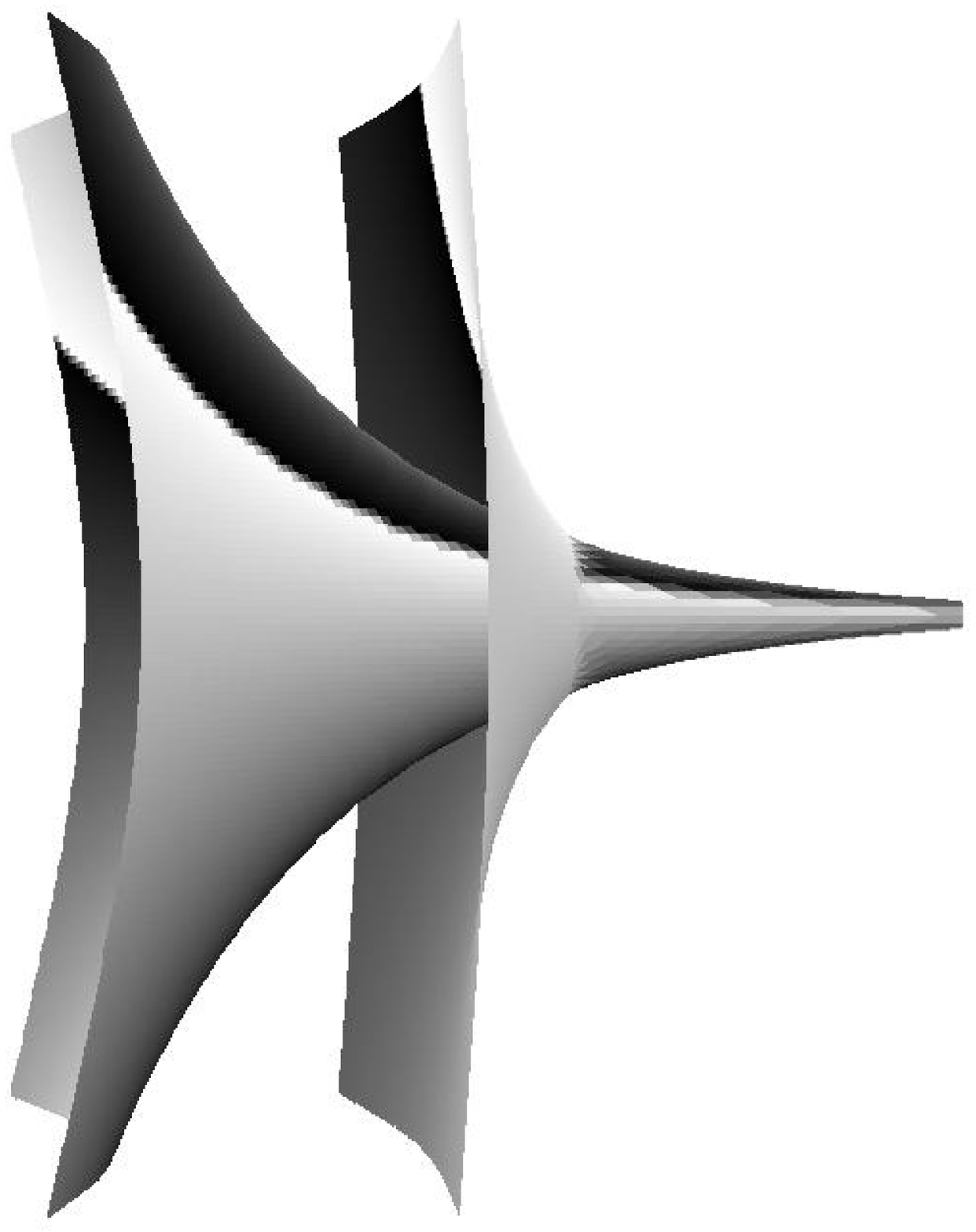}
\end{minipage}%
\begin{minipage}[c]{0.333\textwidth}
{(b) An r=0 vacuum}
\centering
\includegraphics[width=2.2in]{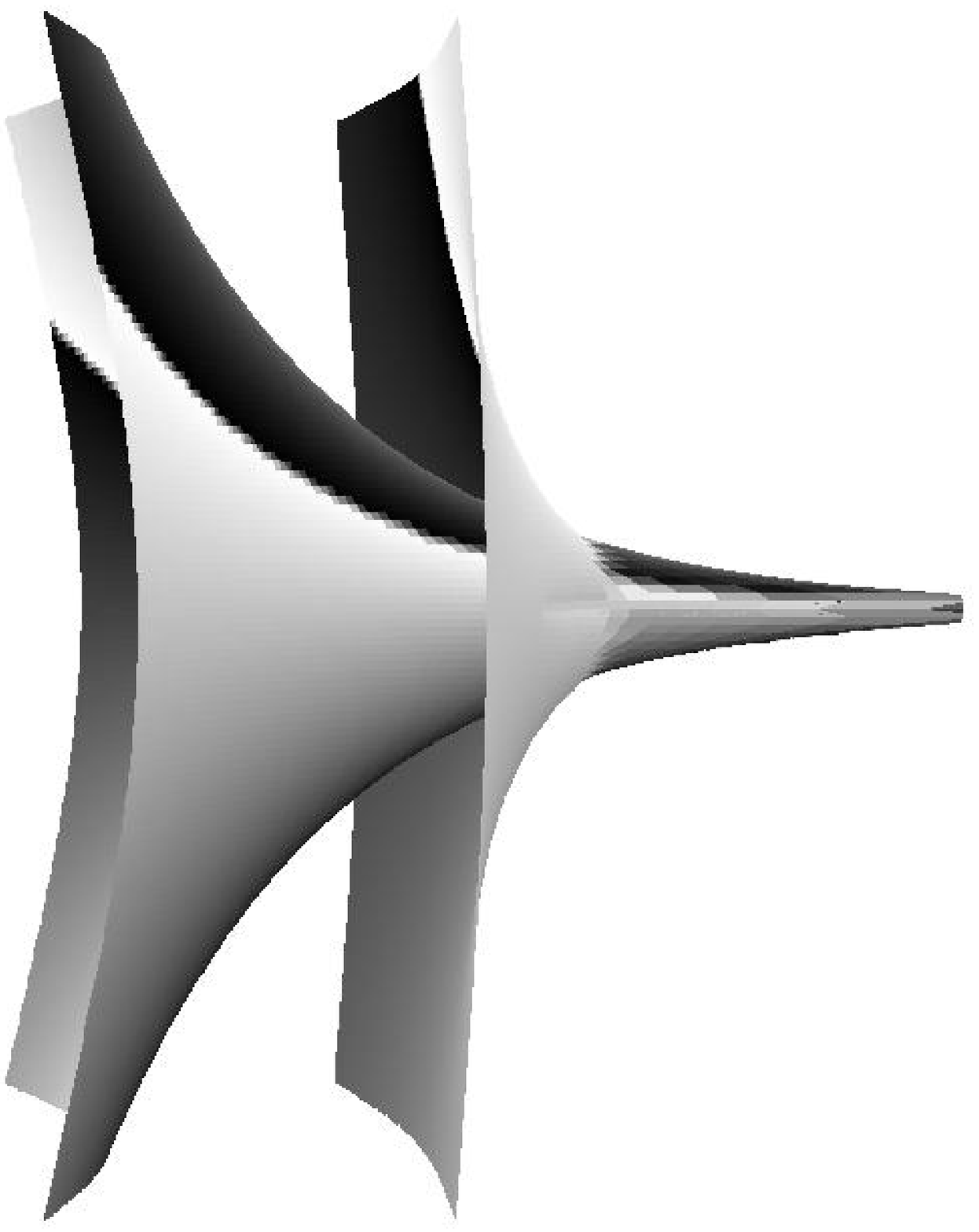}
\end{minipage}%
\begin{minipage}[c]{0.333\textwidth}
{(c) The r=2 vacuum}
\centering
\includegraphics[width=2.2in]{su3nf4r2m=0.eps}
\end{minipage}
\caption{M5 brane configurations at the r vacua in SU(3) with 4 flavors
  and $\forall m_i =0$.}
  \label{m0m5}
\end{figure}

\begin{figure}[htbp]
\begin{center}
\begin{minipage}[c]{.25\textwidth}
\centering
  \includegraphics[width=1.65in]{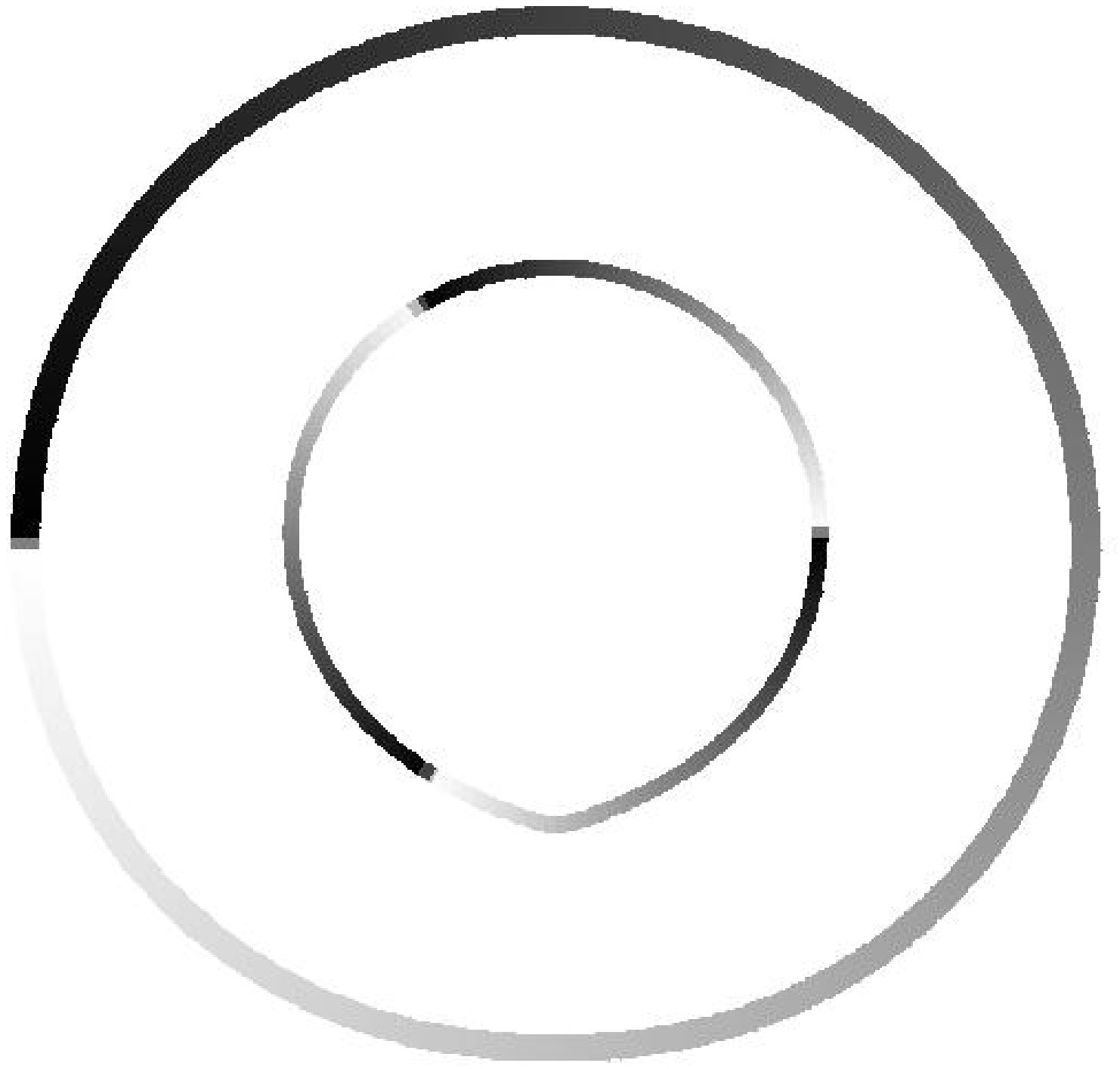}
\end{minipage}%
\begin{minipage}[c]{.25\textwidth}
\centering
  \includegraphics[width=1.45in]{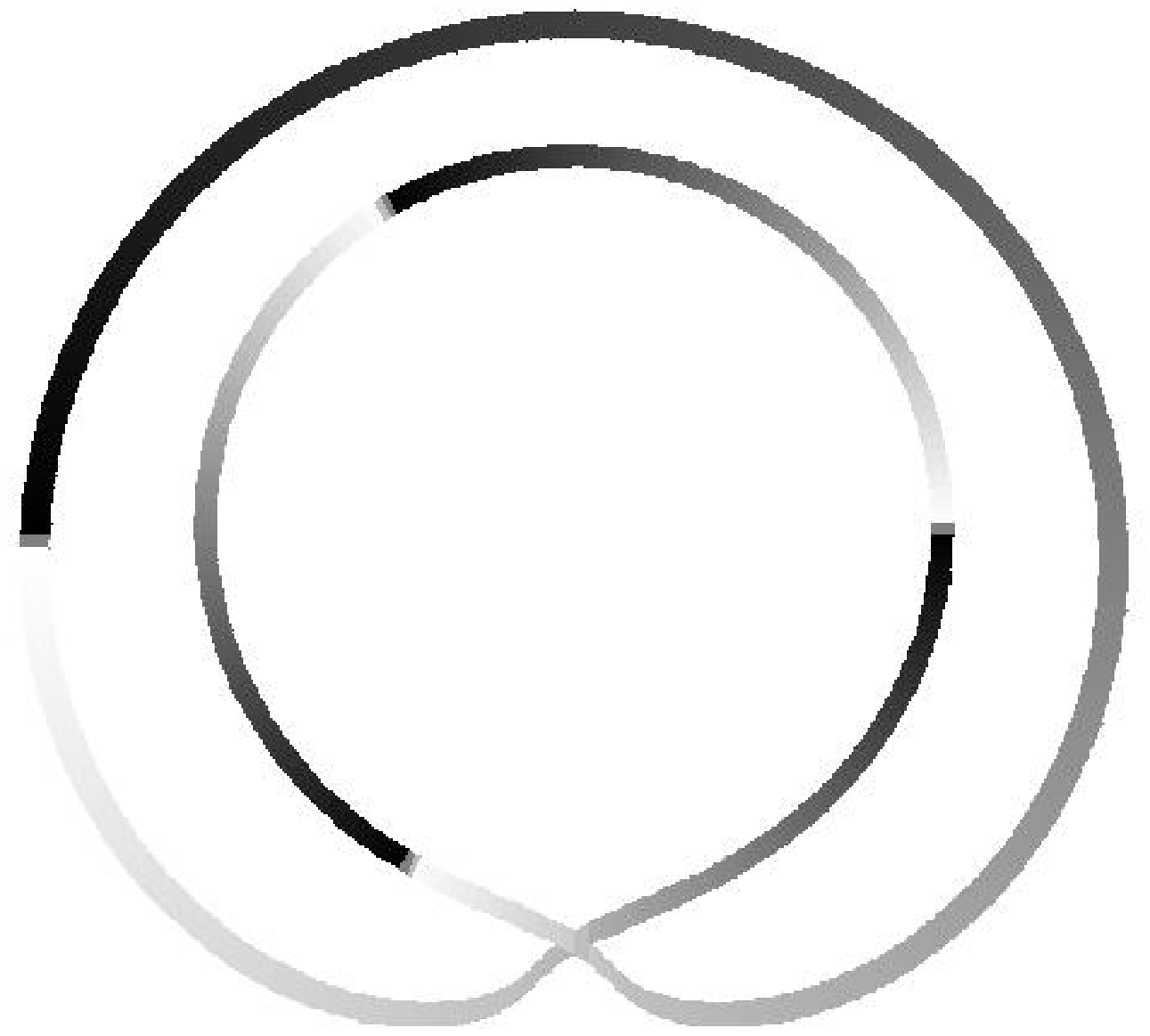}
\end{minipage}%
\begin{minipage}[c]{.25\textwidth}
\centering
  \includegraphics[width=1.35in]{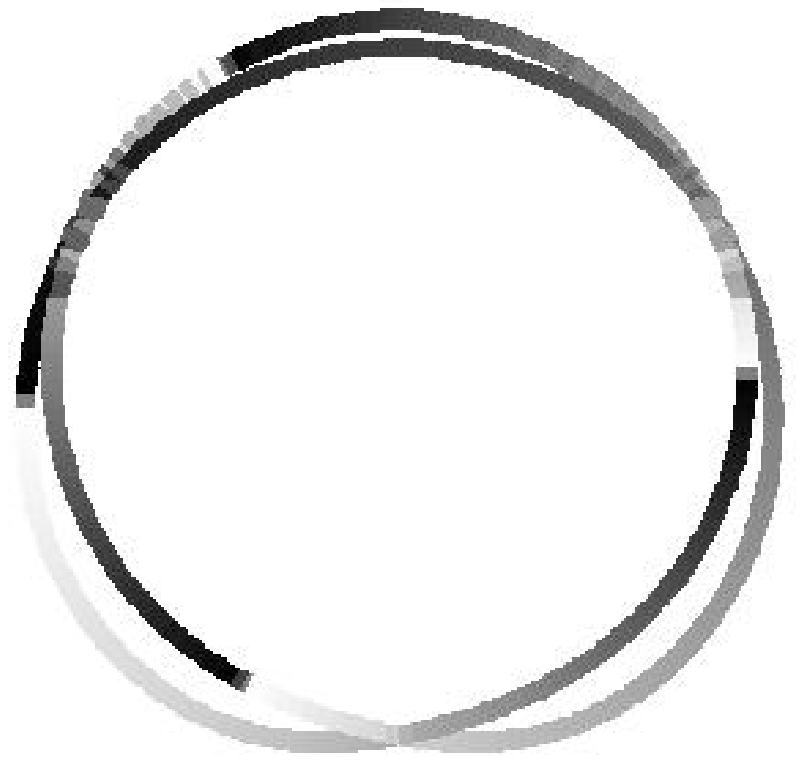}
\end{minipage}
\begin{minipage}[c]{.25\textwidth}
\centering
\includegraphics[origin=c,width=1.25in]{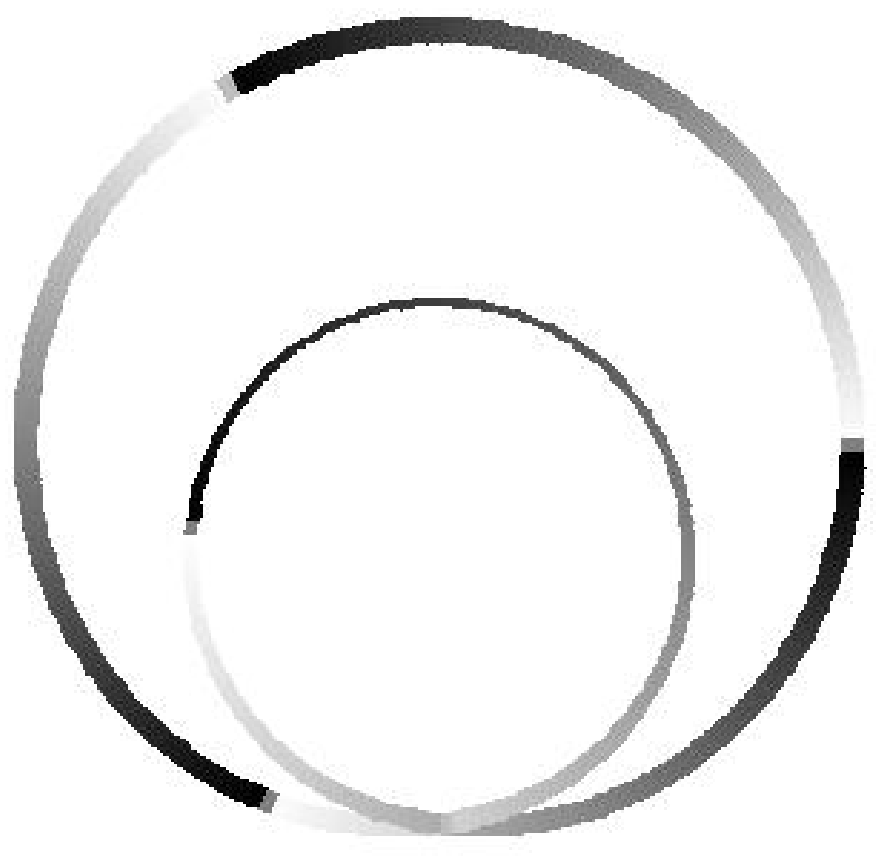}
\end{minipage}%
\begin{minipage}[c]{.25\textwidth}
\centering
  \includegraphics[origin=c,width=0.75in]{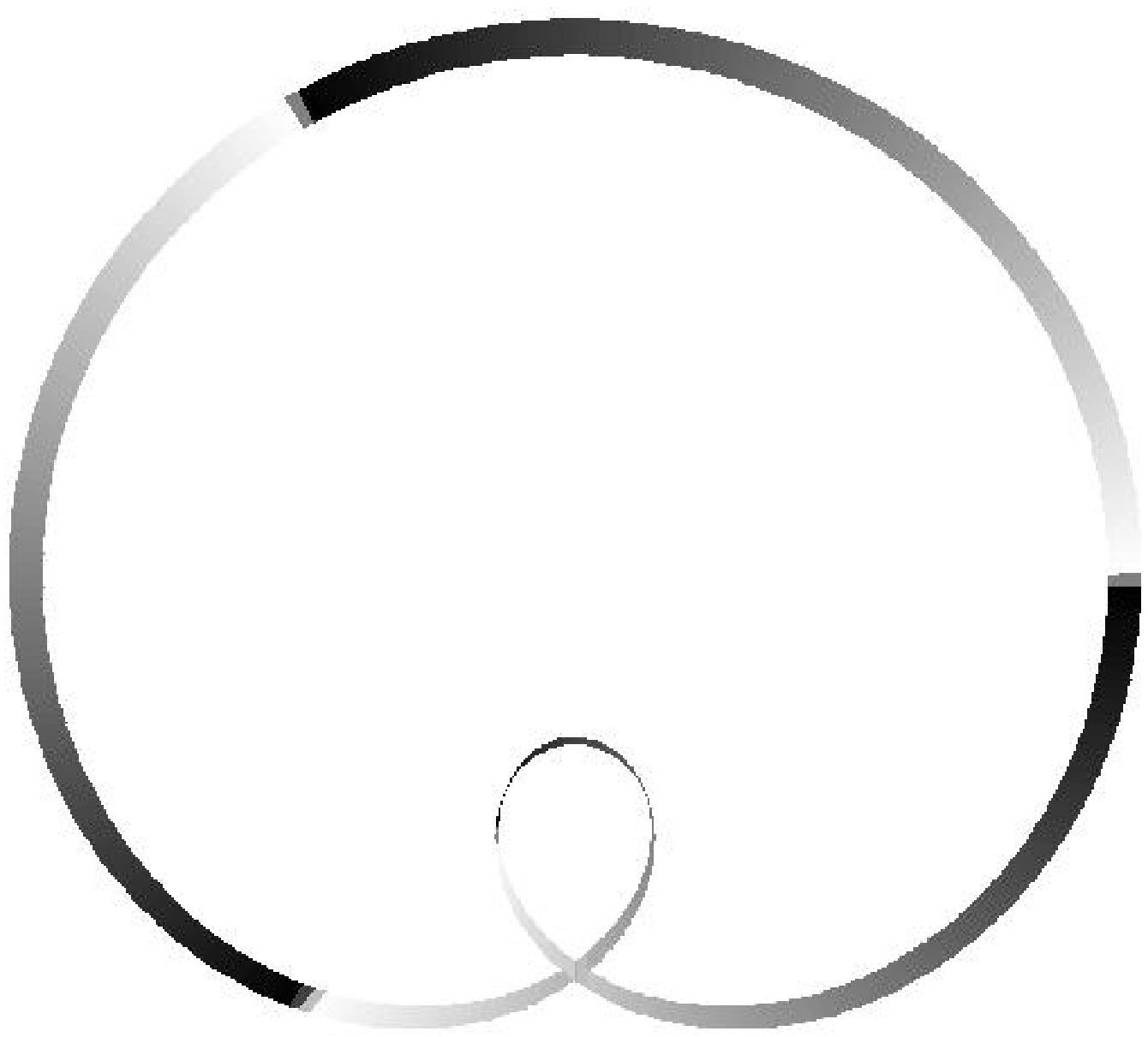}
\end{minipage}%
\begin{minipage}[c]{.25\textwidth}
\centering
  \includegraphics[origin=c,width=1.0in]{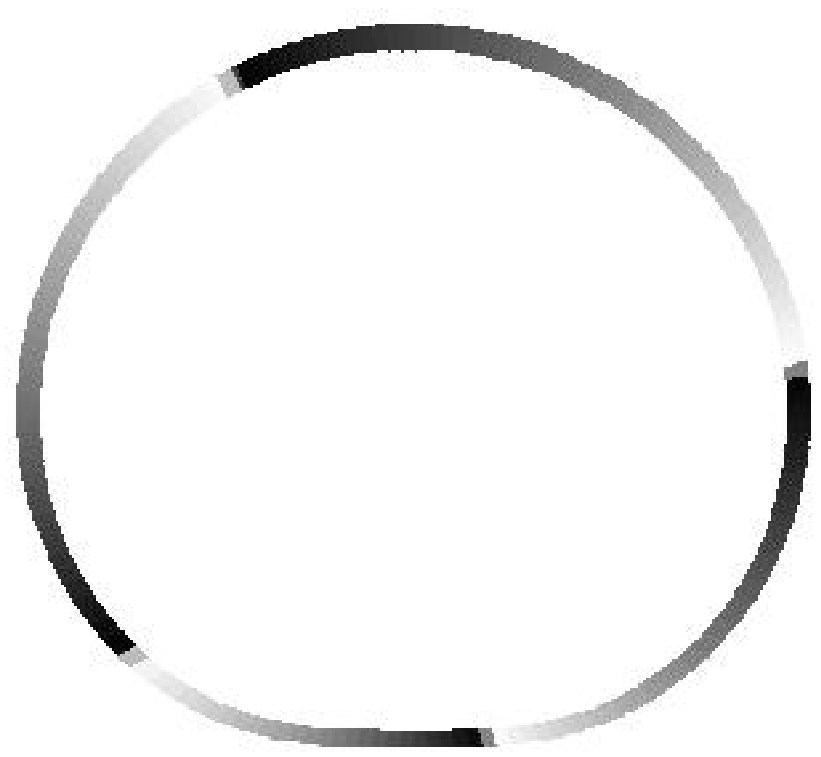}
\end{minipage}
\end{center}
\caption{Cross sections at constant $x^6$ of the M5 brane of Fig.~\ref{m0m5}b (the r=0 root).} \label{m0m5csr0}
\end{figure}

\begin{figure}[htbp]
\begin{center}
\begin{minipage}[c]{.25\textwidth}
\centering
  \includegraphics[width=1.5in]{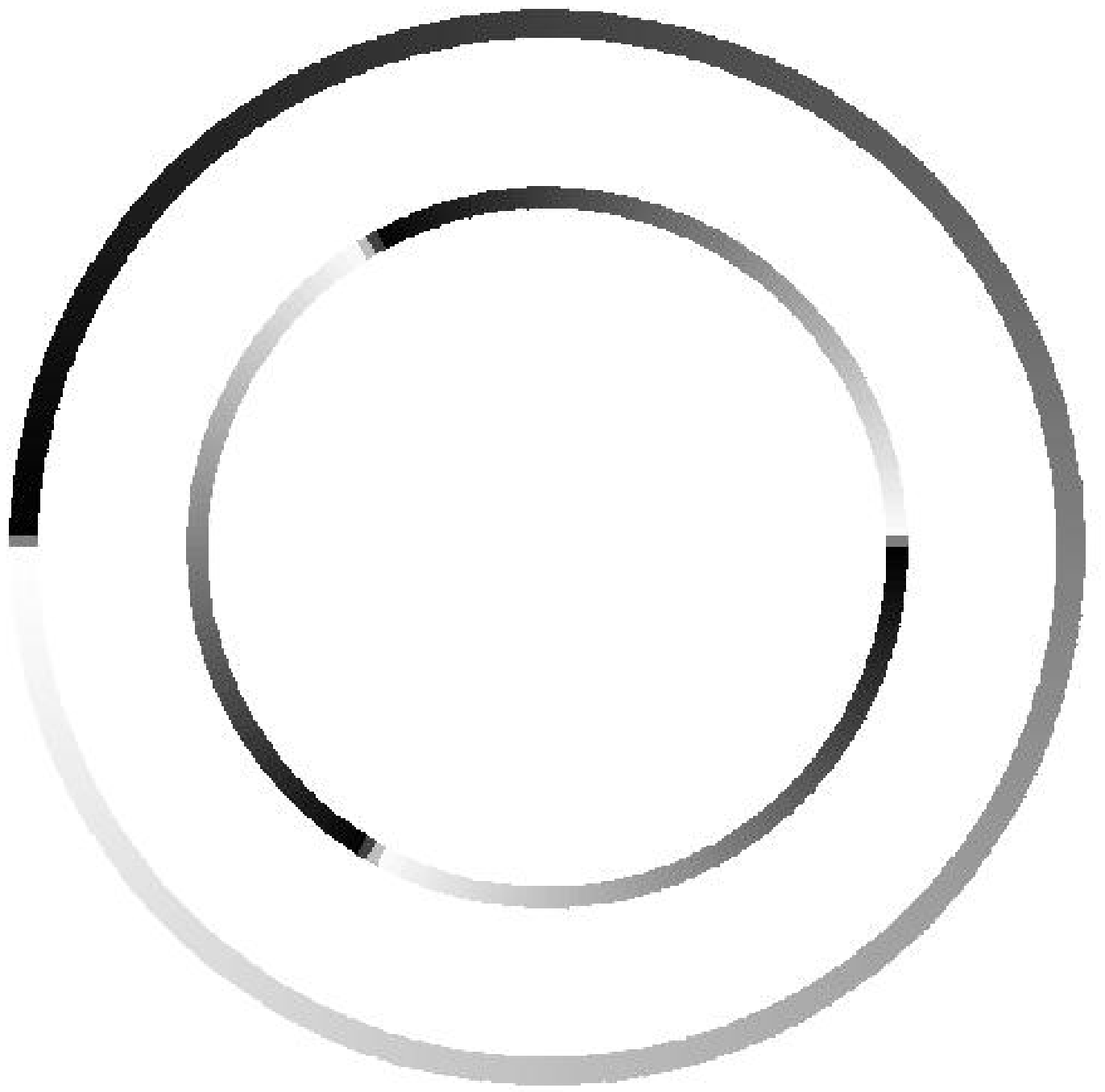}
\end{minipage}%
\begin{minipage}[c]{.25\textwidth}
\centering
  \includegraphics[width=1.5in]{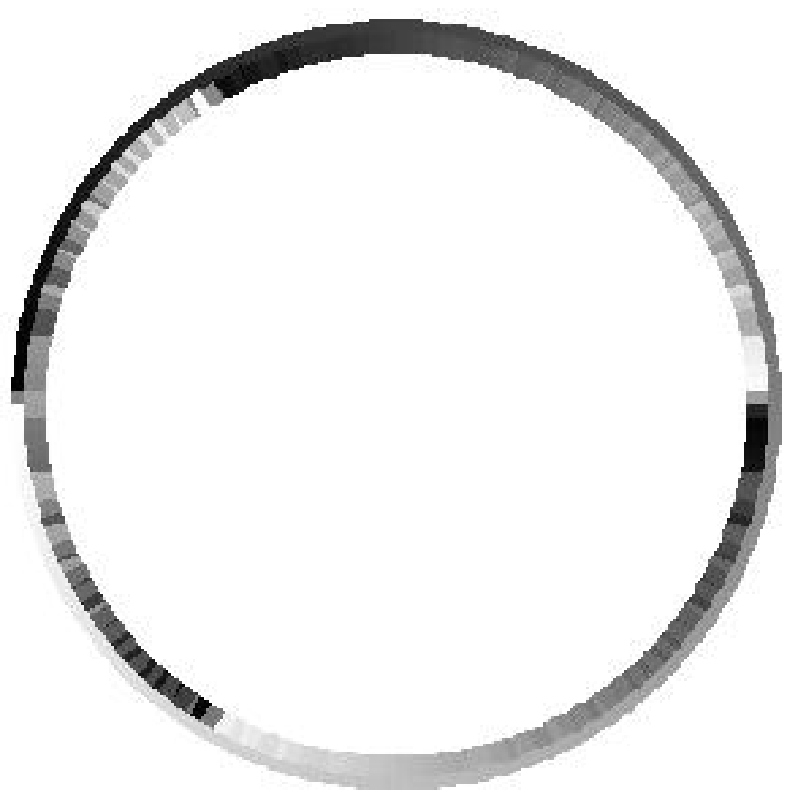}
\end{minipage}%
\begin{minipage}[c]{.25\textwidth}
\centering
  \includegraphics[width=1.5in]{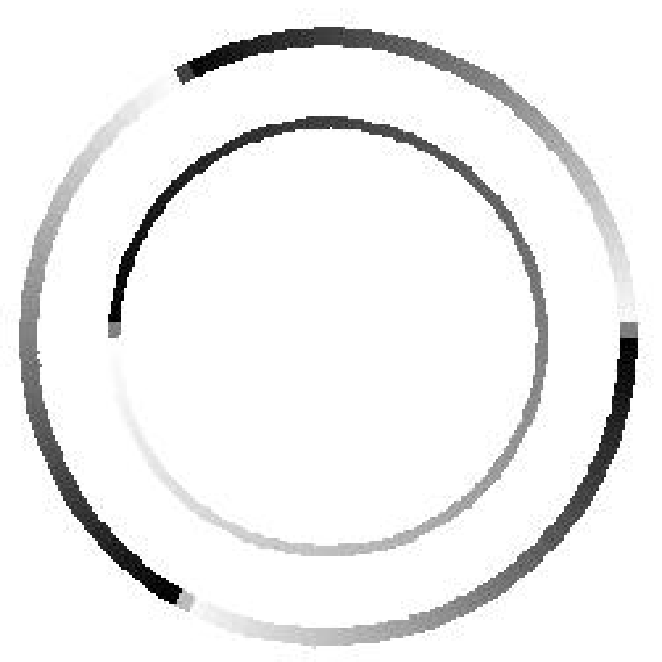}
\end{minipage}%
\begin{minipage}[c]{.25\textwidth}
\centering
  \includegraphics[width=1.5in]{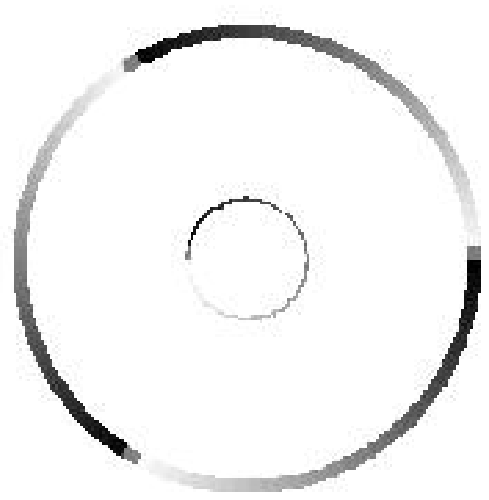}
\end{minipage}%
\end{center}
\caption{Cross sections at constant $x^6$ of the M5 brane of
  Fig.~\ref{m0m5}a (the baryonic root).} \label{m0m5csbr}
\end{figure}

In the case of equal and large bare quark masses, we can clearly see
that the M5 brane configurations which survive breaking to $\N=1$ in
Fig.~\ref{semicm5} do indeed correspond to $r$ flavor branes and color
branes connected to each other.
In the massless case, the relevant curves are more difficult to
interpret, though we include them for completeness in Fig.~\ref{m0m5}.
Note that in the limit that $m \rightarrow 0$, the $r=1$ non-baryonic
root and the baryonic root converge to the same point in moduli space,
the $m=0$ baryonic root.  Just as in the superconformal case, it is
easier to interpret these brane configuration if we consider their
cross-sections in Figs.~\ref{m0m5csr0} and \ref{m0m5csbr}, though we
will leave the interpretation of these cross-sections for future work.

\bibliographystyle{ieeetr} 
\bibliography{m5}
\end{document}